\documentclass[fleqn,usenatbib]{mnras}

\usepackage[normalem]{ulem}

\usepackage[T1]{fontenc}
\usepackage{ae,aecompl}
\usepackage{enumitem}
\usepackage{multirow}
\usepackage{tabularx}

\DeclareRobustCommand{\VAN}[3]{#2}
\let\VANthebibliography\thebibliography
\def\thebibliography{\DeclareRobustCommand{\VAN}[3]{##3}\VANthebibliography}

\usepackage{graphicx}	
\usepackage{amsmath}	
\usepackage{amssymb}	
\usepackage{newtxtext,newtxmath}
\usepackage{fontawesome}
\usepackage{hyperref}

\newcommand{\hi}{H\textsc{i}\ }
\newcommand{\hinospace}{\textrm{H\textsc{i}}}

\newcommand{\secref}[1]{\hyperref[#1]{Section~\ref*{#1}}}
\newcommand{\appref}[1]{\hyperref[#1]{Appendix~\ref*{#1}}}

\title[GPR for HI IM]{Gaussian Process Regression for foreground removal in HI intensity mapping experiments}

\author[P. S. Soares et al.]{Paula S. Soares,$^{1}$\thanks{E-mail: p.s.soares@qmul.ac.uk}
Catherine A. Watkinson,$^{1}$
Steven Cunnington$^{1}$
and Alkistis Pourtsidou$^{1,2}$ 
\\
$^{1}$School of Physics and Astronomy, Queen Mary University of London, Mile End Road, London E1 4NS, UK\\
$^{2}$Department of Physics and Astronomy, University of the Western Cape, Cape Town 7535, South Africa
}

\date{Accepted XXX. Received YYY; in original form ZZZ}

\pubyear{2021}

\begin{document}
\label{firstpage}
\pagerange{\pageref{firstpage}--\pageref{lastpage}}
\maketitle

\begin{abstract}
We apply for the first time Gaussian Process Regression (GPR) as a foreground removal technique in the context of single-dish, low redshift \hi intensity mapping, and present an open-source \texttt{python} toolkit for doing so. We use MeerKAT and SKA1-MID-like simulations of 21cm foregrounds (including polarisation leakage), \hi cosmological signal and instrumental noise. We find that it is possible to use GPR as a foreground removal technique in this context, and that it is better suited in some cases to recover the \hi power spectrum than Principal Component Analysis (PCA), especially on small scales. GPR is especially good at recovering the radial power spectrum, outperforming PCA when considering the full bandwidth of our data. Both methods are worse at recovering the transverse power spectrum, since they rely on frequency-only covariance information. When halving our data along frequency, we find that GPR performs better in the low frequency range, where foregrounds are brighter. It performs worse than PCA when frequency channels are missing, to emulate RFI flagging. We conclude that GPR is an excellent foreground removal option for the case of single-dish, low redshift \hi intensity mapping in the absence of missing frequency channels. Our \texttt{python} toolkit \texttt{gpr4im} and the data used in this analysis are publicly available on GitHub.
\end{abstract}




\section{Introduction}

Neutral hydrogen (\hinospace) intensity mapping (IM) is a novel probe able to trace the large-scale structure (LSS) of the Universe using the 21cm hyperfine transition of hydrogen \citep{Bharadwaj_2001, Battye2004, Chang_2008}. After reionization, most of the neutral hydrogen in the interstellar medium became ionised, and most of the remaining neutral hydrogen can be found inside of self-shielding galaxies. Hence, \hi is a good probe for the distribution of matter in the Universe. Instead of using \hi to trace individual galaxies, IM uses the combined emission of \hi from numerous galaxies to probe how much structure is in a given area of the sky. As such, it can map large areas of the sky very quickly, posing an advantage over traditional spectroscopic galaxy surveys. See \cite{kovetz2017lineintensity} and \cite{Liu_2020} for reviews on the subject.

For single-dish IM \citep{Battye_2013}, several telescope dishes can be used in autocorrelation mode, probing large areas of the sky very quickly \citep{Bull_2015, santos2017meerklass, wang2020hi}. While fast, single-dish IM suffers from low angular resolution, since any structure smaller than the angular resolution of the telescope will not be resolved. This has the effect of damping small scale structure. Despite this, studies have found that it is possible to probe cosmological parameters to significant precision with future \hi IM single-dish surveys (e.g. \citealt{Chang_2008, Peterson:2009ka, Seo:2009fq, Ansari:2012vn, Battye_2013, Bull_2015, Villaescusa_Navarro_2016,Pourtsidou:2016dzn, Soares_2021,kennedy2021statistical}).

In order to access the cosmological \hi IM signal, systematic and instrumental effects must be addressed, of which astrophysical foregrounds pose a particular problem. These foregrounds emit in the same radio frequencies at which we observe the \hi signal, and are several orders of magnitude brighter. However, they vary slowly in frequency (i.e. their spectrum looks very smooth), while the \hi signal is tracing the large-scale structure of the Universe and so is not smooth in frequency.

We can exploit the spectral smoothness of the foregrounds to separate them from the cosmological signal \citep{chang2010, Liu_2011, wolz2015foreground, Alonso_2014fgs, Bigot_Sazy_2015, Olivari_2015, Switzer_2015}. Blind foreground removal methods such as Principal Component Analysis (PCA) do not require previous knowledge of what the foregrounds look like, using the fact that they should be statistically separable from our \hi signal (for a study into some of the main foreground removal methods for \hi IM, see \citealt{cunnington202021cm}). However, it is difficult to avoid also removing some of the \hi signal in the process, leading to biased cosmological parameter estimates \citep{wolz2015foreground, Cunnington_2020nongauss, Soares_2021}.

In reality, foreground signals are not perfectly smooth in frequency, since telescope effects such as polarisation leakage can introduce structure that is difficult to mitigate, posing further complications to foreground removal \citep{Jeli_2008, jelic2010, Moore_2013, Liao_2016, Carucci_2020}. While it is possible to try to simulate polarisation leakage effects, it is important to keep in mind that, in the absence of end-to-end simulations, current foreground simulations are considered overly idealised. Real data analyses of \hi IM experiments conduct much more aggressive foreground cleaning than simulations require. In fact, an auto-correlation \hi IM detection is yet to be observed, with current detections relying on cross-correlation with optical galaxy surveys. Since optical galaxy surveys have different systematic effects than \hi IM surveys, these drop out in cross-correlation, making a detection achievable \citep{Masui:2012zc, Switzer_2013, Wolz_2016, Anderson_2018, wolz2021hi}.

In the realm of higher redshift ($z>6$), interferometric \hi experiments, the aim is to probe the Epoch of Reionization (EoR) and Cosmic Dawn, learning about the first stars and the Universe before reionization. While some instrumental and systematic effects differ between this and our low redshift, single-dish \hi IM case, both are subject to astrophysical foreground contamination. Gaussian Process Regression (GPR; \citealt{carledwardrasmussen2005}) has been used as a foreground removal technique in the context of EoR (\citealt{Mertens_2018}; hereafter M18), and has been found to work well for both simulations and real data \citep{Gehlot_2019, Offringa_2019, Mertens_2020, Ghosh_2020, Hothi_2020, kern2020gaussian}. GPR, like other foreground removal methods such as PCA, can also lead to \hi signal loss. Recent studies have looked at how to correct for this signal loss within GPR \citep{Mertens_2020, kern2020gaussian}, and in this paper we will consider the method proposed in \cite{Mertens_2020}.

In this paper, we aim to apply GPR for the first time as a foreground removal technique in the context of single-dish, low redshift \hi IM. Our analysis follows on from \cite{cunnington202021cm} (hereafter C21), which used MeerKAT and SKA1-MID-like \hi IM simulations, including \hi cosmological signal, smooth foregrounds, polarised foregrounds, and instrumental noise, to thoroughly investigate and compare different foreground removal techniques. We use these simulations to determine how GPR would perform in our case, and compare it to the widely used foreground removal technique PCA. 

We aim to investigate first if it is possible to perform GPR as a foreground removal technique in our case, and if yes, we aim to explore the following questions: How well does it perform compared to PCA? What are its advantages? What are its limitations? How does it perform in the context of different instrumental and systematic effects? What should we keep in mind, and be careful about, if we want to use it in the context of real data analyses?

We base our work on the publicly available code \texttt{ps$\_$eor}\footnote{\href{https://gitlab.com/flomertens/ps_eor}{gitlab.com/flomertens/ps$\_$eor}} (M18), which outlines how to perform GPR in the context of EoR observations. Using it as a great starting point, we created the publicly available \texttt{python} package \texttt{gpr4im}\footnote{\href{https://github.com/paulassoares/gpr4im}{github.com/paulassoares/gpr4im}}, a user-friendly code that can be used to perform GPR as a foreground removal technique for any real space \hi intensity map, either simulated or real. The symbol \faicon{github-alt} in the caption of each figure links to a \texttt{jupyter} notebook showing how the figure was produced.

The paper is organised as follows: We introduce our simulations in \secref{sec:sims}. We explain how GPR and PCA work to remove foregrounds in \secref{sec:fgmethods}. In \secref{sec:results} we show power spectrum results of the recovered \hi power spectrum using GPR and PCA, for the different cases considered. We conclude in \secref{sec:conclusion}.

\section{Simulations}\label{sec:sims}

\begin{figure*}
	\centering
	\includegraphics[width=2\columnwidth]{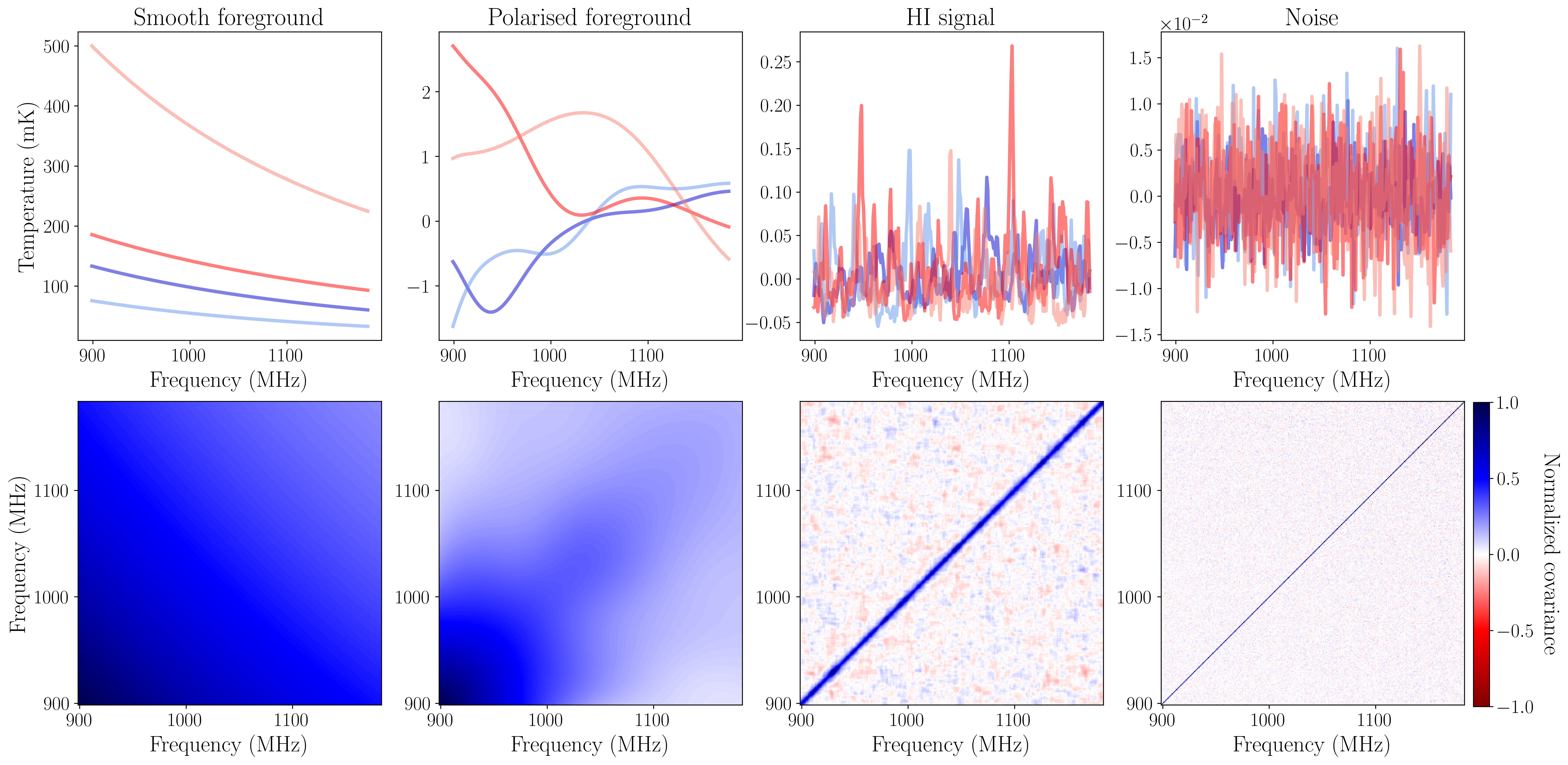}
    \caption{\textit{Top}: Random LoS samples drawn from each component of our simulation, namely from left to right: the smooth foregrounds, the polarised foregrounds, the \hi (21cm) signal, and the instrumental noise. \textit{Bottom}: Normalised frequency covariance for each component. The panels are arranged in order of decreasing signal amplitude and frequency correlation, from left to right. \href{https://github.com/paulassoares/gpr4im/blob/main/Jupyter\%20Notebooks/Reproducible\%20paper\%20plots/Assorted\%20plots.ipynb}{\faicon{github-alt}}}
    \label{fig:data_samples}
\end{figure*}

We make use of the \hi IM simulations presented in C21, choosing the ones centred on the Stripe82 field, a popular target area for galaxy surveys.  These simulations include the smooth foreground signal, the polarised foreground signal, the \hi cosmological signal and instrumental noise. The polarisation leakage is simulated using the software \texttt{CRIME}\footnote{\href{http://intensitymapping.physics.ox.ac.uk/CRIME.html}{intensitymapping.physics.ox.ac.uk/CRIME.html}}, and the smooth foregrounds are extrapolated from observations, which contain sky structure. For more details on the simulations, see \appref{SimAppendix} and C21. They are an idealised representation of what a future \hi IM survey will look like, such as the proposed MeerKLASS survey \citep{santos2017meerklass}. We assume throughout that our sky area is small enough that we can use a Cartesian grid projection without much distortion. This flat-sky approximation is widely used in large-scale structure surveys, where curved sky effects are not a major limitation \citep{Blake_2018, Castorina2018}.

Each frequency channel ($\nu$) of our simulation is binned into pixels ($\boldsymbol{\theta}$), therefore the total observed temperature fluctuations in our data can be described as:
\begin{equation}
    \delta T_{\rm obs}(\nu, \boldsymbol{\theta}) = \delta T_{\rm FG}(\nu, \boldsymbol{\theta}) + \delta T_{\rm HI}(\nu, \boldsymbol{\theta}) + \delta T_{\rm noise}(\nu, \boldsymbol{\theta}) \, . 
\end{equation}

The foreground signal is itself composed of the Galactic synchrotron emission, free-free emission, extragalactic point sources and polarisation leakage:
\begin{equation}
    \delta T_{\rm FG} = \delta T_{\rm sync} + \delta T_{\rm free} + \delta T_{\rm point} + \delta T_{\rm pol} \, .
\end{equation}

We centre our simulations at an effective redshift of $z_{\rm eff} = 0.39$, and assume a redshift range of $0.2 < z < 0.58$, though our cosmological signal does not evolve with redshift (see \appref{SimAppendix}). The frequency at which we observe the \hi signal changes with redshift as $\nu = 1420\text{MHz}/(1+z)$, so our redshift range corresponds to a frequency range of $899 < \nu < 1184 \, \text{MHz}$, which is representative of the MeerKAT telescope in the L-band \citep{santos2017meerklass}. The dimensions of our simulaion box are, in comoving distance units, $L_\text{x} = 1000,\, L_\text{y} = 1000,\, L_\text{z} = 924.78\,\, \text{Mpc}\,h^{-1}$, and we grid it into volume-pixels (voxels) of dimensions $N_\text{x} = 256,\, N_\text{y} = 256,\, N_\text{z} = 285$ (z denotes the parallel to the line-of-sight (LoS) direction, while x and y are perpendicular to the LoS). This gridding corresponds to a frequency resolution of $\delta \nu = 1\text{MHz}$. The sky area covered by our simulation is approximately 3000 deg$^2$, also representative of a MeerKLASS-like survey \citep{santos2017meerklass}.

After adding our different signals together, we smooth them by a constant Gaussian telescope beam, to make the angular resolution representative of future surveys. The resolution of the beam is determined by the frequency of observation ($\nu$) and the size of the telescope dish ($D_{\rm dish}$):
\begin{equation}
    \theta_{\rm FWHM} = \frac{1.22c}{\nu D_{\rm dish}} \, ,
\end{equation}
where $c$ is the speed of light. We choose $D_{\rm dish} = 15$ m based on the SKA1-MID dishes, but this is also similar to the MeerKAT dish size. Although $\theta_{\rm FWHM}$ is frequency dependent, we smooth every frequency channel to the same resolution to aid the performance of our foreground removal algorithms \citep{matshawule2020hi}. We choose the common resolution to be that at our minimum frequency, $\nu_{\rm min} = 899\,\text{MHz}$, since this will give the largest beam size, equivalent to $\theta_{\rm FWHM} = 1.55$ deg.

The top row of \autoref{fig:data_samples} shows each component of our simulation as a function of frequency. Each line shows a different random pixel, to give a better idea of what the signal looks like throughout our simulation box. The foreground signals are much smoother in frequency than the \hi signal, and this will be used later in order to perform the foreground removal. The polarised foreground is less smooth in frequency than the smooth foregrounds, which will be important for foreground removal too. On the bottom row, we show the normalised frequency covariance of each component of our signal. It will be relevant later that the smooth foreground signal is brighter and more correlated at lower frequencies.

\section{Foreground removal methods}\label{sec:fgmethods}

Here we describe the main foreground removal technique studied in this paper, GPR. We also present the PCA technique, which is a widely used foreground removal technique that we will compare the performance of GPR to.

\subsection{Gaussian Process Regression}

In this section, we will introduce the framework of GPR and discuss how it applies to our foreground removal problem. For a comprehensive and in depth discussion of Gaussian processes please see \cite{carledwardrasmussen2005}, and for an intuitive description see \cite{luger2021mapping}. We used these, as well as M18, \cite{Mertens_2020} and \cite{kern2020gaussian}, as excellent references for writing this section.

We can model our simulated data as a vector $\textbf{d}$, where each element is one ``observation'' (while our data is simulated, we will use the term ``observation'' in this paper for simplicity). The length of this data vector is determined by the number of observations made (in our case, $N_{\nu} = N_{\rm z} = 285$ observations made at $N_{\nu}$ frequency channels). Each element (or observation) is an image \textit{slice} with total number of pixels given by $N_\theta = N_{\rm x} \times N_{\rm y}$, and containing all the spatial information observed at that frequency. Therefore, our data vector is actually a data \textit{matrix} with dimensions $N_\theta \times N_{\nu}$. The frequency range of our data, a vector with length $N_{\nu}$, is denoted as $\boldsymbol{\nu}$.

Our data is composed of four signal components: the smooth foregrounds ($\textbf{f}_{\rm smooth}$), the polarised foregrounds ($\textbf{f}_{\rm pol}$), the \hi cosmological signal ($\textbf{f}_{\rm 21}$) and the instrumental (Gaussian) noise ($\textbf{n}$). We write our data matrix as the sum of these components:
\begin{equation}
    \textbf{d} = \textbf{f}_{\rm fg} + \textbf{f}_{\rm 21} + \textbf{n} \, ,
\end{equation}
where our foregrounds $\textbf{f}_{\rm fg}$ can be made up of either only a smooth component ($\textbf{f}_{\rm fg} = \textbf{f}_{\rm smooth}$), or a smooth component and a polarised component ($\textbf{f}_{\rm fg} = \textbf{f}_{\rm smooth} + \textbf{f}_{\rm pol}$). The smooth component is smooth in frequency, while the polarised one is less so, as seen in \autoref{fig:data_samples}.

Because we are assuming that our signal components are separate from each other, the (frequency) covariance ($\textbf{C}$) of our data ($\textbf{d}$) can be written as
\begin{equation}\label{datacov}
\textbf{C} = \langle \textbf{d}\textbf{d}^{\rm T} \rangle = \textbf{d}\textbf{d}^{\rm T}/(N_\nu - 1) = \textbf{C}_{\rm fg} + \textbf{C}_{\rm 21} + \textbf{C}_{\rm n} \, ,
\end{equation}
where the foreground covariance is written as $\textbf{C}_{\rm fg} = \textbf{C}_{\rm smooth}$ in the absence of polarisation, and $\textbf{C}_{\rm fg} = \textbf{C}_{\rm smooth} + \textbf{C}_{\rm pol}$ in the presence of polarisation. Here we are assuming that the polarised foregrounds have an additive effect, which is a suitable approximation to first order. Formally, due to mode-mixing, polarisation leakage would also have a multiplicative effect \citep{Mertens_2020}.

\subsubsection{What is a Gaussian process?}

The main idea behind GPR is that we can describe each component of our data \textbf{d} as a \textit{Gaussian process}. A Gaussian process is a Gaussian distribution over infinite dimensions. Think of a multivariate Gaussian distribution over two variables, but extend the number of variables to infinity. And instead of variables, it is functions of frequency $\boldsymbol{\nu}$. Each line in the top panel of \autoref{fig:data_samples} is a function, representing signal in our data ($\textbf{d}$) which is a function of frequency ($\boldsymbol{\nu}$).

To understand this better, think of how a multivariate Gaussian distribution over variables is defined by a mean $\boldsymbol{\mu}$ (of size $N \times 1$) and covariance $\boldsymbol{\Sigma}$ (of size $N \times N$). A random variable $f$ ($N \times 1$) that has a multivariate Gaussian distribution defined by $\boldsymbol{\mu}$ and $\boldsymbol{\Sigma}$ is written as
\begin{equation}
    f \sim \mathcal{N}(\boldsymbol{\mu}, \boldsymbol{\Sigma}) \, .
\end{equation}

Now, assume that the mean is actually a \textit{mean function}, in our case a function of frequency: $\boldsymbol{\mu} = m(\boldsymbol{\nu}) \equiv m$. The only difference is that now each element of the mean depends on frequency: $\mu_i = m(\nu_i)$. Assume also that the covariance is defined by a \textit{kernel function}: $\text{K} \equiv k(\boldsymbol{\nu}, \boldsymbol{\nu})$, so each element of the covariance is given by $\boldsymbol{\Sigma}_{i,j} = k(\nu_i, \nu_j)$. If we assume a random variable $f$ is a Gaussian process, we write
\begin{equation}\label{gpform}
\begin{split}
    \ & f \sim \mathcal{GP}(m,\text{K}) \,, \\
    \ & f(\boldsymbol{\nu}) \sim \mathcal{N}(m(\boldsymbol{\nu}), k(\boldsymbol{\nu}, \boldsymbol{\nu})) \, .
\end{split}
\end{equation}

Multivariate Gaussian distributions have several useful properties, many of which are shared by Gaussian processes. We outline a few of these below:

\begin{itemize}[leftmargin=*]
    \item Multivariate Gaussian distribution (and Gaussian processes) are easy to sample from. All one needs is a mean (or mean function) and a covariance (or kernel function);
    \item Multivariate Gaussian distributions (and Gaussian processes) have a well defined marginal likelihood function, assuming only Gaussian noise is present. It can be calculated analytically given a mean (or mean function) and a covariance (or kernel function), and can be maximised for inference problems;
    \item A Gaussian process is a \textit{stochastic} process. Since we drawing from an infinite Gaussian distribution, each draw is random and can take any functional form consistent with the mean and kernel function.
\end{itemize}

We assume that our data \textbf{d} (and each of its components) is one such variable $f$, which is drawn from a Gaussian process with a particular mean and kernel function as in \autoref{gpform}. As is commonly done with GPR, we assume a zero mean function, which is true for our data since we mean-centre each frequency slice.

\subsubsection{Foreground removal with GPR}

If we consider our data \textbf{d} to be a Gaussian process defined over the frequency range $\boldsymbol{\nu}$, we can obtain its joint probability with a Gaussian process defined at another set of frequencies $\boldsymbol{\nu'}$ as
\begin{equation}
    \begin{bmatrix}
    \textbf{d} \\
    \textbf{d}'
    \end{bmatrix}
    = \mathcal{N} \left(
    \begin{bmatrix}
    0 \\
    0
    \end{bmatrix}, 
    \begin{bmatrix}
    k(\boldsymbol{\nu}, \boldsymbol{\nu}) & k(\boldsymbol{\nu}, \boldsymbol{\nu'}) \\
   k(\boldsymbol{\nu'}, \boldsymbol{\nu}) & k(\boldsymbol{\nu'}, \boldsymbol{\nu'})
    \end{bmatrix}
    \right) \, .
\end{equation}

Since our data is made up of separate components (each of which is a Gaussian process), we split its kernel function ($\text{K} \equiv k(\boldsymbol{\nu}, \boldsymbol{\nu})$) into separate components:
\begin{equation}
    \text{K} = \text{K}_{\rm fg} + \text{K}_{\rm 21} + \text{K}_{\rm n} \, ,
\end{equation}
where if we have polarised foregrounds presents, we write $\text{K}_{\rm fg} = \text{K}_{\rm smooth} + \text{K}_{\rm pol}$. Otherwise, $\text{K}_{\rm fg} = \text{K}_{\rm smooth}$.

We are interested in separating our foreground signal from the rest of our data. If we know our data's kernel function $\text{K}$, we can write the joint probability distribution of our data \textbf{d} and the foreground model $\textbf{f}_{\rm fg}$ as (M18):
\begin{equation}
    \begin{bmatrix}
    \textbf{d} \\
    \textbf{f}_{\rm fg}
    \end{bmatrix}
    = \mathcal{N} \left(
    \begin{bmatrix}
    0 \\
    0
    \end{bmatrix}, 
    \begin{bmatrix}
    \text{K}_{\rm fg} + \text{K}_{\rm 21} + \text{K}_{\rm n} & \text{K}_{\rm fg} \\
    \text{K}_{\rm fg} & \text{K}_{\rm fg}
    \end{bmatrix}
    \right) \, 
\end{equation}

Our foreground model $\textbf{f}_{\rm fg}$ is also a Gaussian process, with expectation value ($\text{E}[\textbf{f}_{\rm fg}]$) and covariance ($\text{cov}[\textbf{f}_{\rm fg}]$) defined by
\begin{equation}
    \textbf{f}_{\rm fg} \sim \mathcal{N}(\text{E}[\textbf{f}_{\rm fg}], \text{cov}[\textbf{f}_{\rm fg}]) \, .
\end{equation}

The expectation value is a prediction of the foreground signal, and the covariance is the uncertainty in this prediction. These are given by (M18):
\begin{equation}
    \begin{split}
        \ & \text{E}[\textbf{f}_{\rm fg}] = \text{K}_{\rm fg}[\text{K}_{\rm fg} + \text{K}_{\rm 21} + \text{K}_{\rm n}]^{-1}\textbf{d} \, , \\
        \ & \text{cov}[\textbf{f}_{\rm fg}] = \text{K}_{\rm fg} - \text{K}_{\rm fg}[\text{K}_{\rm fg} + \text{K}_{\rm 21} + \text{K}_{\rm n}]^{-1}\text{K}_{\rm fg}
    \end{split}
\end{equation}

\begin{figure*}
	\centering
	\includegraphics[width=2\columnwidth]{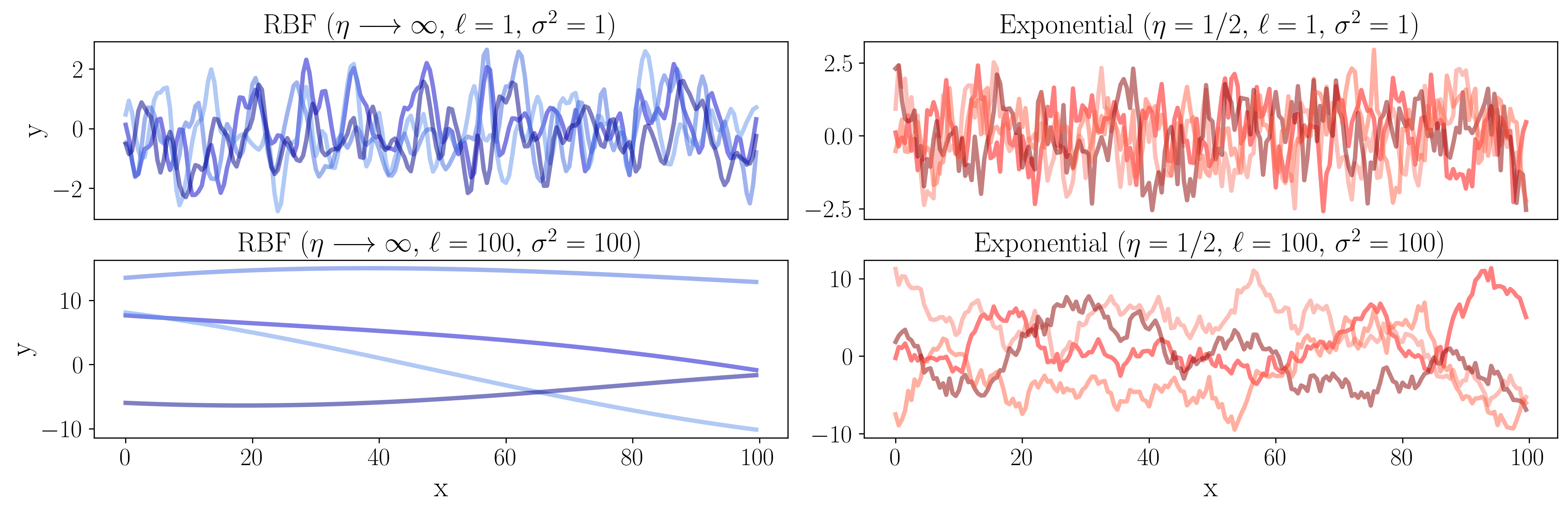}
    \caption{\textit{Left}: RBF covariance samples. \textit{Right}: Exponential covariance samples. We can see that overall, the RBF samples look smoother than the exponential samples. The top row shows the case of small variance and small lengthscale ($\ell = 1,\, \sigma^2=1$), meaning the signal amplitude is small and the data is not very correlated in frequency. The bottom row is showing the opposite case of a higher variance and lengthscale ($\ell = 100,\, \sigma^2=100$), with higher signal amplitude and more frequency correlation. \href{https://github.com/paulassoares/gpr4im/blob/main/Jupyter\%20Notebooks/Reproducible\%20paper\%20plots/Assorted\%20plots.ipynb}{\faicon{github-alt}}}
    \label{fig:samples_len_var}
\end{figure*}

The expectation value $\text{E}[\textbf{f}_{\rm fg}]$ is dependent not only on the foreground kernel $\text{K}_{\rm fg}$ but also on the full kernel of our data $\text{K} = \text{K}_{\rm fg} + \text{K}_{\rm 21} + \text{K}_{\rm n}$. It is thus important to ensure we have an appropriate kernel for all the elements in our data. $\text{E}[\textbf{f}_{\rm fg}]$ is also dependent on our data \textbf{d}, so it can be thought of as: given the known data \textbf{d}, the foreground kernel $\text{K}_{\rm fg}$ and the full data kernel $\text{K}$, what is the expected (mean) value of our foregrounds in our data's frequency range?

Thus, by estimating $\textbf{f}_{\rm fg}$, we are making an informed prediction for what our foregrounds look like. We can then subtract this prediction from the data, and hopefully be left with only our cosmological signal and instrumental noise. This is how one performs foreground removal with GPR. The residual vector \textbf{r} (what is leftover after GPR foreground removal) is defined as (M18):
\begin{equation}
    \textbf{r} = \textbf{d} - \text{E}[\textbf{f}_{\rm fg}] \, .
\end{equation}

We are also interested in $\text{cov}[\textbf{f}_{\rm fg}]$, the ``error margin'' of our foreground prediction $\text{E}[\textbf{f}_{\rm fg}]$. This is important for bias correction techniques (see \secref{sec:bias}).

So far we have described the process of foreground removal with GPR assuming we know the kernel $\text{K} = \text{K}_{\rm fg} + \text{K}_{\rm 21} + \text{K}_{\rm n}$. In reality, we need to use the data to try and find the best-fitting kernel for each signal component.

\subsubsection{Kernel functions}

To understand how we find the optimised kernel functions for our data, let's first think of a simple linear regression problem. Consider a vector \textbf{y} of observed values (dependent variable), and a vector \textbf{x} of input values (independent variable), both of length $N$. If the variables follow a linear trend, it can be fit with a linear model: $y_i = mx_i + b, \, i=1,...,N$. This might be a good model for the data, but the model parameters ($m$ and $b$) still need to be optimised, which can be done using a maximum likelihood estimation, for example. Thus, this is a two-step process: first, a model is chosen to best describe the data, and then the parameters of the model are optimised.

Our kernel optimisation problem is similar, but not as simple as this two-step process. We wish to find the kernel functions K that best fit the covariance of our data \textbf{C} (\autoref{datacov}). For example, for the foreground covariance $\textbf{C}_{\rm fg}$, we want to find the kernel function $\text{K}_{\rm fg}$ that best describes that covariance, and its best-fitting hyperparameters. The parameters of the kernel functions are called hyperparameters to emphasise that a  Gaussian process is a non-parametric model.

We henceforth use the terms ``kernel'', ``kernel function'' and ``covariance function'' interchangeably, but the \textit{data} covariance is not equivalent to the covariance \textit{function} - the covariance function is the model we are optimising in order to best describe the data covariance.

The covariance function must be a function of frequency ($\text{K} = k(\boldsymbol{\nu}, \boldsymbol{\nu})$), since our data is a function of frequency. There exist several widely used kernels in GPR. One of them is the Mat{\'e}rn kernel, which has the form \citep{carledwardrasmussen2005}:
\begin{equation}
    \text{K}_{\rm Matern}(\nu, \nu') = \sigma^2 \frac{2^{1-\eta}}{\Gamma(\eta)} \left( \sqrt{2\eta} \frac{| \nu - \nu' |}{\ell} \right)^{\eta} K_\eta \left( \sqrt{2\eta} \frac{| \nu - \nu' |}{\ell} \right) \, ,
\end{equation}
where $\Gamma$ is the gamma function and $K_\eta$ is the modified Bessel function of the second kind. The three hyperparameters of our Mat{\'e}rn kernel model that we want to optimise are: 
\begin{itemize}[leftmargin=*]
    \item $\sigma^2$ is the \textit{variance}, which describes the overall amplitude of the signal. We expect $\sigma^2$ to be larger for our foregrounds than for our \hi signal since the foregrounds are brighter (see \autoref{fig:data_samples});
    \item $\ell$ is the \textit{lengthscale}, which describes the typical scale of correlations in our data, across frequency. A larger value of $\ell$ means the data is more correlated in frequency, so we expect $\ell$ to be larger for our foregrounds than for our \hi signal, since the foregrounds are more correlated in frequency (see \autoref{fig:data_samples});
    \item $\eta$ is the \textit{spectral parameter}, which determines the overall ``smoothness'' of our data. We expect this to be larger for our foregrounds than for our \hi signal, since our foregrounds look smooth in frequency while the \hi signal is more Gaussian (see \autoref{fig:data_samples}).
\end{itemize}

The spectral parameter $\eta$ is particularly important because it can simplify the Mat{\'e}rn kernel. Some common choices are:
\begin{itemize}[leftmargin=*]
    \item $\eta \rightarrow \infty$: in this limit, the Mat{\'e}rn kernel simplifies to a radial basis function (RBF), also called a squared exponential function or Gaussian function. This limit describes a very smooth signal, which is ideal for our foreground signal. Both \cite{kern2020gaussian} and \cite{Mertens_2020} use this kernel to describe their smooth foreground component. It has the form:
    \begin{equation}\label{rbfkern}
        \text{K}_{\rm RBF}(\nu, \nu') = \sigma^2 \exp \left( \frac{( \nu - \nu' )^{2}}{2\ell^{2}} \right)\, ;
    \end{equation}
    \item $\eta = 5/2$: in this limit, the Mat{\'e}rn kernel is relatively smooth but not as smooth as the RBF kernel. M18 uses this to describe their smooth foreground component;
    \item $\eta = 3/2$: in this limit, the Mat{\'e}rn kernel is even less smooth in frequency, and it can be used to describe foreground components that have medium frequency smoothness, as is done in e.g. M18 and \cite{Mertens_2020};
    \item $\eta = 1/2$: in this limit, the Mat{\'e}rn kernel is the least smooth in frequency, and is known as the exponential function. It agrees well with a spectrally varying signal such as the \hi signal. The form of this kernel, which is used in e.g. M18, \cite{Mertens_2020} and \cite{kern2020gaussian} to describe the \hi signal, is:
    \begin{equation}\label{expkern}
        \text{K}_{\rm exp}(\nu, \nu') = \sigma^2 \exp \left( \frac{| \nu - \nu' |}{\ell} \right)\, .
    \end{equation}
\end{itemize}

The Mat{\'e}rn kernel is very useful and widely applicable due to its flexibility. With different hyperparameters, it can describe a variety of different signals. We find that the Mat{\'e}rn kernel is sufficient to describe all components in our data, with differing hyperparameters. Other kernel functions, such as the rational quadratic function, are also useful. \cite{Gehlot_2019} for example use it to describe frequency varying foregrounds, which are qualitatively similar to our polarised foregrounds, however we did not find it to perform as well as the Mat{\'e}rn kernel in our data.

There also exist kernel functions with periodic behaviour, for more oscillatory signals. Our simulations do not include this type of behaviour hence we do not consider these kernels. However, with real data, there might exist systematic effects with this type of behaviour.

In \autoref{fig:samples_len_var}, we exemplify what samples drawn from an exponential and RBF covariance look like. We consider the cases of small variance and lengthscale ($\ell = 1,\, \sigma^2=1$), as well as a larger variance and lengthscale ($\ell = 100,\, \sigma^2=100$) to illustrate the difference. In both cases, the RBF samples are much smoother than the exponential samples, due to it having a higher spectral parameter ($\eta \rightarrow \infty$). For both the RBF and exponential kernels, increasing the variance $\sigma^2$ leads to an increase in the amplitude of the signal, and increasing the lengthscale $\ell$ makes the samples more correlated in frequency.

\begin{figure}
	\centering
	\includegraphics[width=\columnwidth]{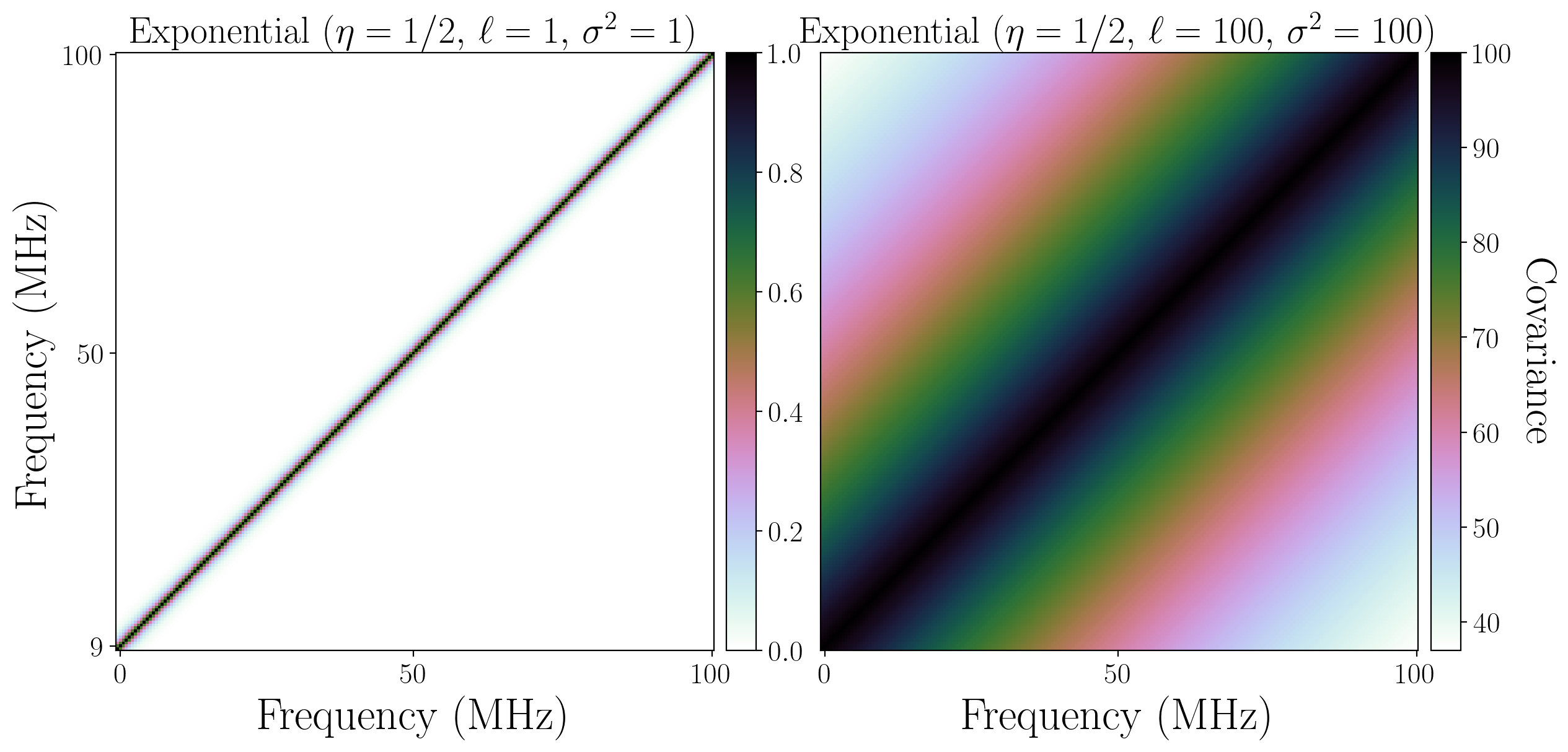}
    \caption{Exponential kernel for two different cases. \textit{Left}: Low variance, low lengthscale case ($\ell = 1,\, \sigma^2=1$). \textit{Right}: High variance, high lengthscale case ($\ell = 100,\, \sigma^2=100$). \href{https://github.com/paulassoares/gpr4im/blob/main/Jupyter\%20Notebooks/Reproducible\%20paper\%20plots/Assorted\%20plots.ipynb}{\faicon{github-alt}}}
    \label{fig:cov_example}
\end{figure}

We also demonstrate what the exponential kernel function looks like. In \autoref{fig:cov_example}, we show the exponential kernel in two different cases: the low variance, low lengthscale case ($\ell = 1,\, \sigma^2=1$), and the high variance, high lengthgscale case ($\ell = 100,\, \sigma^2=100$). Increasing the variance changes the amplitude of the covariance, as seen in the colour bar, while increasing the lengthscale changes the correlation of the signal, as seen by how the diagonal correlation is much wider in the second plot. A wider diagonal means that the data is more correlated in frequency, while a smaller diagonal means that the data points are mostly correlated with themselves and those immediately around them.

\subsubsection{Hyperparameter optimisation and model selection}\label{sec:modelselec}

We have established that the kernel hyperparameters affect the behaviour of the samples being drawn from it. How do we simultaneously pick the best kernel model (e.g. choice of $\eta$ for a Mat{\'e}rn kernel), and also the best hyperparameters (e.g. lengthscale and variance)? These steps are interdependent. When comparing different models, we want to compare the best overall performance of each model. Therefore, we first optimise the hyperparameters of our model by maximising the marginal likelihood over the function parameters (which are distinct from the model hyperparameters defined above); we then compare its evidence (i.e. the marginal likelihood over the function parameters) to other optimised models. When we talk about ``model'' in this section, we are talking about finding the kernel function choices that best model our data (what we have referred to as K in previous sections).

Given a model (kernel function) with hyperparameters $\boldsymbol{\theta}$, the marginal likelihood of the data (also known as the Bayesian evidence) can be calculated as the integral of the data likelihood $\mathcal{L}$ times the prior:
\begin{equation}\label{evidence}
    p(\textbf{d} | \boldsymbol{\nu}, \boldsymbol{\theta}) = \int \mathcal{L}(\textbf{d}|\textbf{x}, \boldsymbol{\nu}, \boldsymbol{\theta})p(\textbf{x}|\boldsymbol{\nu}, \boldsymbol{\theta})d\textbf{x} \, ,
\end{equation}
where \textbf{x} is another vector occupying the same space as our data, which describe the functions of the Gaussian process. The result of this integration is a constant, called the log-marginal likelihood (LML):
\begin{equation}\label{lml}
    \log p(\textbf{d} | \boldsymbol{\nu}, \boldsymbol{\theta}) = -\frac{1}{2} \textbf{d}^{\rm T}\text{K}^{-1}\textbf{d} -\frac{1}{2} \log |\text{K}| - \frac{n}{2}\log 2\pi \, ,
\end{equation}
where $n$ is the number of data points sampled, and K is again the full kernel function for the data (the sum of the kernels for each component of the data, including noise). We assume that each component of our data is a Gaussian process, and that the noise is Gaussian distributed, thus the LML can be calculated analytically. The LML not only favours models that fit our data well - it also disfavours overly complex models \citep{carledwardrasmussen2005}.

The LML tells us the ``evidence'' of the model, i.e. how well our covariance function fits our data. We can take the ratio of two evidences to compare how well different models fit our data. This is known as the Bayes factor $\mathcal{Z}$. When working with the log evidence, the log of the Bayes factor is achieved by taking the difference between the LML of two different models (called m1 and m2):
\begin{equation}
    \log \mathcal{Z} = \text{LML}_{\rm m1} - \text{LML}_{\rm m2} \, ,
\end{equation}
where larger values of $\mathcal{Z}$ or $\log \mathcal{Z}$ indicate that m1 fits the data better than m2.

We are also interested in obtaining the posterior distribution of our hyperparameters. According to Bayes' theorem, we can obtain the posterior probability density by multiplying the likelihood with the prior. In log space, the log posterior probability density can be written as:
\begin{equation}
    \log p(\boldsymbol{\theta}|\textbf{d},\boldsymbol{\nu}) \propto \log p(\textbf{d}|\boldsymbol{\nu}, \boldsymbol{\theta}) + \log p(\boldsymbol{\theta})\, .
\end{equation}

Because the LML is easy to calculate with \autoref{lml}, we can calculate it for several choices of hyperparameters $\boldsymbol{\theta}$ and find the combination of $\boldsymbol{\theta}$ that gives us the \textit{maximum likelihood} - the largest value for LML. This method for maximising the LML is known as Type-II maximum likelihood (ML-II, e.g. gradient descent), and yields point estimates for the hyperparameters. While fast, ML-II has several caveats, as it is prone to: overfitting when there are many hyperparameters, getting stuck in local minima, and underestimating predictive uncertainty by yielding only point estimates for the best-fit values of the hyperparameters (without any uncertainty) \citep{simpson2020marginalised}. Some ML-II methods, such as simulated annealing or ensemble sampling, could lead to better global maxima estimates, but are still unable to provide uncertainties on the hyperparameters. We use the \texttt{python} package \texttt{GPy}\footnote{\href{https://github.com/SheffieldML/GPy}{github.com/SheffieldML/GPy}} \citep{gpy2014} to run gradient descent in the context of GPR.

Nested sampling (NS) is a technique able to sample complex distributions, and it focuses on estimating the model evidence (\autoref{evidence}) and well as its uncertainty. It also yields the posterior distribution of the hyperparameters as a by-product, which is useful. We used the \texttt{python} package \texttt{pymultinest}\footnote{\href{http://johannesbuchner.github.io/PyMultiNest/}{johannesbuchner.github.io/PyMultiNest}} \citep{Buchner_2014, Feroz_2009} to run NS.

A downside to this method is that it is computationally expensive. We compared our model evidence estimate from NS to that of gradient descent, and found them to be consistent. We therefore use \texttt{GPy} for quickly comparing different models, and NS for obtaining the final best-fitting hyperparameter results of our chosen model. Our model was relatively simple, with few hyperparameters, so both gradient descent and NS were able to sample the hyperparameter space easily. For more complex models or real data, the hyperparameter space will likely be more complex, leading to differences between gradient descent and NS, so it would be better to use NS in practice. Our simulations are idealised and do not suffer from systematic effects other than polarisation leakage, thus it was straightforward to find the best-fitting models since the differences between the evidences of different kernel choices were very large (much larger than the offset between the NS and gradient descent evidences).

We summarise our model selection and hyperparameter optimisation steps below:
\begin{enumerate}[leftmargin=*]
    \item Choose a kernel model K, and optimise it using gradient descent, obtaining an estimate for the evidence;
    \item Compare the evidence of the model with that of other models (e.g. try a different kernel for the smooth foregrounds $\text{K}_{\rm fg}$);
    \item Conclude the model selection by using the Bayes factor to find the best model;
    \item Once a model has been selected, run NS to obtain a more robust estimate of the evidence and hyperparameters;
    \item If the posterior distributions from NS look sensible, take the peaks of the distributions as the final optimised hyperparameter values. Though not done in this analysis, it is also possible to take into account the error on this prediction by looking at the standard deviation of the posterior distributions.
\end{enumerate}

During model selection, one should carefully monitor the best-fit hyperparameter estimates to check if the result is hitting against a prior or seems unphysical given what is known about the data. For example, we would not expect the variance of the \hi kernel to be orders of magnitude larger than the variance of the foreground kernel - we would expect the opposite. If this is happening, the model is not a good fit for the data, regardless of the evidence.

\begin{figure}
	\centering
	\includegraphics[width=\columnwidth]{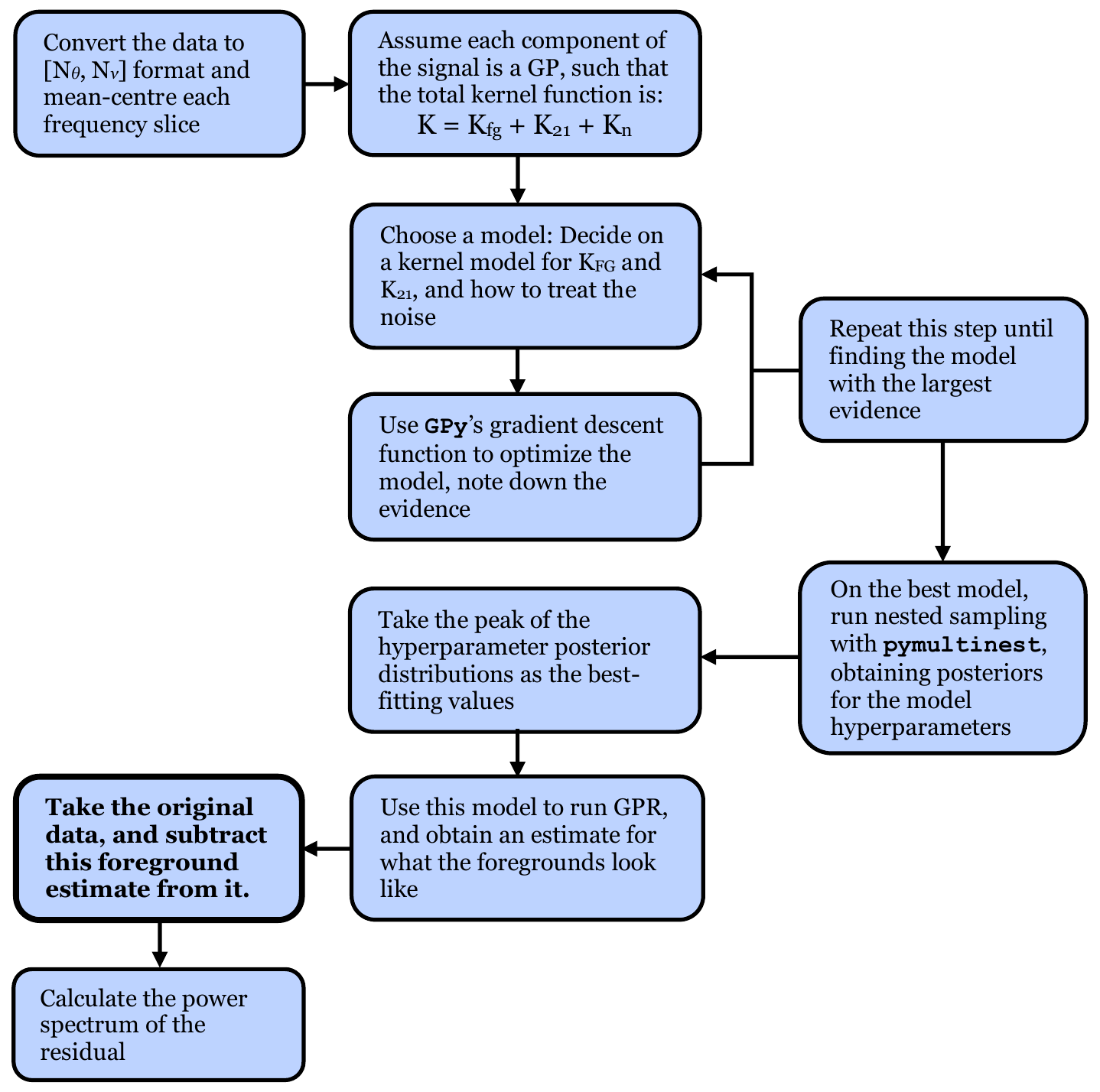}
    \caption{A flowchart outlining our foreground removal pipeline, starting with our data in [$N_{\rm x}$,$N_{\rm y}$,$N_{\rm z}$] format, and ending up with the power spectrum estimate of our foreground cleaned residual. In bold is the step where foreground removal is performed.}
    \label{fig:flowchart}
\end{figure}

\subsubsection{Noise treatment}\label{gprnoise}

There are a few different ways to treat Gaussian noise in GPR. We note some below:

\begin{enumerate}[leftmargin=*]
    \item Constant noise: Assume the data has some random Gaussian noise that is constant in frequency, with variance given by $\sigma^2_n$. GPR will then assume that the observed target value $y$ is given at a particular frequency by
    \begin{equation}
        y = f(\boldsymbol{\nu}) + \epsilon \,,
    \end{equation}
    where $\epsilon \sim \mathcal{N}(0, \sigma^2_n)$, and $f$ is defined in \autoref{gpform}. We can think of this as adding a new kernel K$_n$ to our model K, given by K$_n(\nu, \nu') = \sigma^2_n \delta_{\nu, \nu'}$ where $\delta_{\nu, \nu'}$ is a Kronecker delta which is one where $\nu = \nu'$ and zero elsewhere. The kernel can also be written as K$_n = \sigma^2_n I$ where $I$ is the identity matrix. If one has a reasonable estimate for the noise in the data, $\sigma^2_n$ can be fixed. Otherwise, it is a free hyperparameter that can be optimised.
    \item Heteroscedastic noise: This is similar to the constant noise case, with one difference: we assume that the Gaussian noise variance is \textit{not} constant in frequency. We would then have that for a given frequency $\nu_i$:
    \begin{equation}
        y_i = f(\nu_i) + \epsilon_i \,
    \end{equation}
    where $\epsilon_i \sim \mathcal{N}(0, \sigma^2_{n,i})$, $\sigma^2_{n,i}$ being the frequency dependent noise variance. One would then write the noise kernel as K$_n(\nu, \nu') = \sigma^2_n(\nu, \nu')$, where $\sigma^2_n(\nu_i, \nu'_i) = \sigma^2_{n,i}$ for $\nu_i = \nu'_i$ and zero otherwise. It is also possible to either fix this noise variance for each frequency, or let them be free hyperparameters. As discussed in \secref{noise}, in our case the noise variance is changing with frequency.
    \item Indistinguishable noise: Assume that there is indistinguishable noise present in the data, such that $y = f(\boldsymbol{\nu})$. This might mean that there is either no noise present in the data, or that there is noise present but it is indistinguishable from other signals, and so the covariance function of other signals is also describing the noise.
    \item Another noise kernel: Choose a kernel to represent the noise signal, such as an exponential kernel.
\end{enumerate}

For our data, we tried each of these methods and used the Bayes factor to decide on the best approach. We find that for our analysis, case (iii) is the best, possibly due to how small the noise is making it indistinguishable from the faint \hi cosmological signal.

In real data, more complex instrumental noise is likely to be present. For example, time correlated 1/$f$ noise is a problem in \hi intensity mapping experiments, and would require more careful treatment, such as its own kernel (see e.g. \citealt{Li_2020}).

We summarise our pipeline for removing foregrounds from our data using GPR and obtaining the residual power spectrum in \autoref{fig:flowchart}.

\subsubsection{Bias correction}\label{sec:bias}

Ideally, we want our estimate of the foreground removed residual \textbf{r} to be unbiased. Unfortunately, for current foreground removal methods this is not the case. Foreground removal methods especially struggle to recover the true \hi power spectrum on large scales, where our desired \hi signal looks similarly correlated in frequency to the foregrounds and is erroneously removed (see e.g. C21).

For GPR, this is also a problem. Previous works have tried to account for over or under-cleaning with GPR by using \textit{bias correction} techniques. \cite{Mertens_2020} presented an additive bias correction to the residual covariance and power spectrum, which arises analytically assuming the data covariance is perfectly known. \cite{kern2020gaussian} showed that this correction fails to recover the true EoR 21cm power spectrum when the EoR signal covariance is misestimated. While the EoR signal is still undetected and not fully understood, the \hi signal from IM follows the well known large-scale structure of the Universe. Therefore, we are more likely to estimate the \hi signal covariance of our data accurately, making this bias correction a better approximation in our case (especially when considering idealised simulations).

\cite{kern2020gaussian} point out that such an additive bias correction by definition cannot account for \textit{under}-cleaning, only over-cleaning of the signal. They derive a multiplicative bias correction as part of their optimal quadratic estimator for the power spectrum, and show that it is robust to imperfect estimations of the EoR signal covariance. Their correction by construction cannot under-predict the 21cm signal. However, it requires perfectly knowing the foreground covariance. 

Since our simulations are idealised and we are confident about our data covariance estimations, we show the effect of the \cite{Mertens_2020} bias correction to our results, and leave comparison between the \cite{Mertens_2020} and \cite{kern2020gaussian} corrections for future work. We only show this correction to demonstrate its effect in our case, and do not consider it when comparing our GPR results with PCA. The bias correction is analogous to the transfer function, commonly applied to PCA cleaned data, in that it is a data-driven approach for estimating signal loss in a foreground clean \citep{Switzer_2015}. Since we do not apply a transfer function correction to PCA, we do not consider the bias correction to GPR when comparing the two. We want to establish their differences before any signal loss correction is applied. Ideally we want the need for correction to be as minimal as possible, hence it is important to compare methods before corrections.

The \cite{Mertens_2020} bias correction is well described in their Section 3.3.2, and we summarise it here.
For the covariance of the residual, $\langle \textbf{r} \textbf{r}^{\rm T} \rangle$, the bias correction is applied by adding the foreground model covariance ($\text{cov}[\textbf{f}_{\rm fg}]$) to the result:
\begin{equation}
    \langle \textbf{r}\textbf{r}^{\rm T} \rangle_{\rm unbiased} = \langle \textbf{r} \textbf{r}^{\rm T} \rangle\, + \text{cov}[\textbf{f}_{\rm fg}] \, .
\end{equation}

For the radial power spectrum, we calculate the bias correction as outlined below.

\begin{enumerate}[leftmargin=*]
    \item Draw a number of random realisations from a multivariate Gaussian distribution with mean zero and covariance given by $\text{cov}[\textbf{f}_{\rm fg}]$. The realisations are drawn in the same dimensions as our data ($N_\theta \times N_\nu$);
    \item Measure the radial power spectrum for each of these realisations;
    \item Take the average of these power spectra;
    \item Add the averaged power spectrum to the residual power spectrum obtained with GPR.
\end{enumerate}

The procedure for calculating the bias correction for the spherically averaged power spectrum is similar. The only difference is that in step (ii), we bin the measured radial power spectrum into the power spectrum $k$-bins. This ensures we are ignoring the transverse $k_\perp$ modes in our spherically-averaged power spectrum. This is essential since these will only contain random Gaussian noise due to how the simulations are generated in step (i). The only physically relevant information in our simulations is in the radial direction since $\text{cov}[\textbf{f}_{\rm fg}]$ only contains radial information. No bias correction is applied to the transverse power spectrum because $\text{cov}[\textbf{f}_{\rm fg}]$ does not give us any information about what uncertainties might be present in the transverse direction.

\subsection{Principal Component Analysis}

Principal Component Analysis (PCA) is a blind component separation method able to remove astrophysical foregrounds from \hi IM data. It does not require prior knowledge of what the foregrounds look like in order to work. Methods similar to PCA have been used in the analysis of real \hi IM data since it does not make many assumptions about the data. This is useful since we do not yet understand the instrumental systematic effects of \hi IM experiments well enough \citep{Masui:2012zc, Wolz_2016, Anderson_2018}. We choose this popular foreground removal method to compare the performance of GPR to.

PCA tries to estimate the foreground signal by transforming the data to a dimensional basis which maximises variance. In this new context, we expect the first few basis vectors - which contain highly correlated signal with the largest amplitude - to represent the foregrounds. These basis vectors are called the \textit{principal components}, and we designate the number of principal components containing the foreground information as $N_{\rm FG}$.

Recall that we can describe our data as a matrix \textbf{d} with dimensions $N_\theta \times N_\nu$. We assume that our data can be represented as a linear system
\begin{equation}
    \textbf{d} = \textbf{A}\textbf{s} + \epsilon \, ,
\end{equation}
where \textbf{A} is the mixing matrix, containing the amplitude of the $N_{\rm FG}$ separable components (\textbf{s}), and $\epsilon$ is the residual signal, which includes the noise and cosmological signal. The residual is what we want to obtain, and can be obtained simply by rearranging the above equation
\begin{equation}
    \epsilon = \textbf{d} - \textbf{A}\textbf{s} \, .
\end{equation}

This foreground removed residual $\epsilon$ is the PCA equivalent to the residual \textbf{r} we obtain with GPR, and we will compare which of these two are closer to the true residual. For more detail of how exactly PCA finds \textbf{A} and \textbf{s} and performs foreground removal, please see C21.

\section{Results}\label{sec:results}

Here we present our main results, which compare the foreground removal performance of GPR to that of PCA. We do this first in the idealised case of no polarisation leakage, then add in the polarisation leakage to see how it affects our results. We also investigate how dependent our results are on bandwidth and redshift, and whether a foreground removal transfer function is possible with GPR.

When comparing GPR and PCA, we always show two choices of $N_{\rm FG}$ for PCA: one that leads to an under-cleaning of the foregrounds, and one that leads to an over-cleaning. This will help us more robustly compare the GPR and PCA performances. We also show the residuals between what GPR or PCA predict versus the true underlying signal power spectrum. We conclude that a technique is working better if it yields residuals that are close to zero, i.e. a prediction that is close to the truth. We point out when methods under- or overpredict the true power spectrum, but make no conclusion about which of these predictions is best. Overpredicting the power spectrum means there are residual foregrounds left in the data, which can bias results and boost errors. Underpredicting is caused by a loss of the desired underlying signal, which also biases results. However, signal loss can be in principle be corrected with transfer functions \citep{Switzer_2015}, but since we do not fully explore this, leaving it to future work, we make no conclusions about which circumstance is more desirable between under- and overprediction of the power spectrum.

Throughout, we also show the power spectrum results including the bias correction discussed in \secref{sec:bias}. This is only to show the effect of this bias correction, and we do not consider it when comparing the performance to PCA which would also require some treatment for signal loss, such as a transfer function, to ensure a fair comparison. 
We use the terms ``spherically averaged power spectrum'' and ``power spectrum'' interchangeably, but specify when we are talking about transverse or radial power spectra. 

We define the largest accessible scale for the spherically averaged power spectrum to be $k_{\rm min} = 2\pi/V^{1/3}$, where $V$ is the total comoving volume of our data. For the radial power spectrum, this becomes $k_{\rm min,\parallel} = 2\pi/L_{\rm z}$, and for the transverse power spectrum $k_{\rm min,\perp} = 2\pi/\sqrt{(L_{\rm x}^2 + L_{\rm y}^2)}$. Throughout, we assume a bin width of twice the smallest accessible wavenumber.

\subsection{No polarisation}

In the no polarisation case, we want to find what kernel best describes the smooth foregrounds and the \hi signal, and choose a way to treat the noise, as discussed in \secref{gprnoise}. Using our model selection pipeline (which includes a Bayes factor analysis, see \secref{sec:modelselec}), we find that the exponential kernel (\autoref{expkern}) best describes the \hi signal, and the RBF kernel (\autoref{rbfkern}) best describes the smooth foregrounds.

\begin{figure}
	\centering
	\includegraphics[width=\columnwidth]{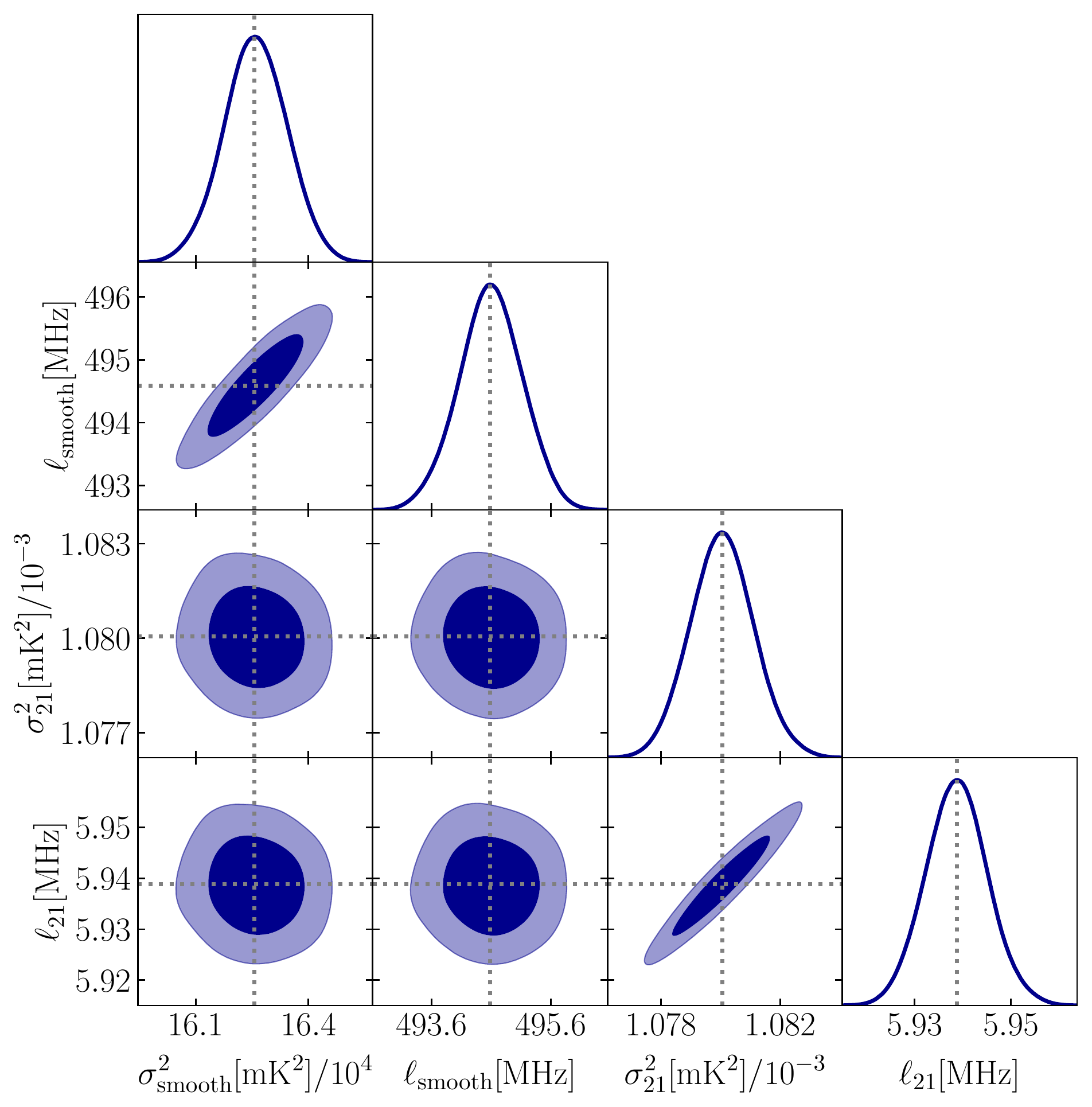}
    \caption{Posterior distribution of our model hyperparameters, obtained using nested sampling in the no polarisation case. Dotted grey lines show the best-fitting hyperparameter results obtained with gradient descent. \href{https://github.com/paulassoares/gpr4im/blob/main/Jupyter\%20Notebooks/Reproducible\%20paper\%20plots/No\%20polarisation.ipynb}{\faicon{github-alt}}}
    \label{fig:posterior_nopol}
\end{figure}

\begin{table*}
	\centering
	\begin{tabular}{ | l | cccccc | } 
        \multicolumn{7}{c}{Median and  1$\sigma$ error estimates for our hyperparameters} \\
        \hline
        Hyperparameter & $\sigma^2_\mathrm{smooth}[\mathrm{mK}^2] / 10^4$ & $\ell_\mathrm{smooth}[\mathrm{MHz}]$ & $\sigma^2_\mathrm{pol}[\mathrm{mK}^2]$ & $\ell_\mathrm{pol}[\mathrm{MHz}]$ & $\sigma^2_{21}[\mathrm{mK}^2]/ 10^{-3}$ & $\ell_{21}[\mathrm{MHz}]$ \\
        \hline\hline
		No polarisation & 16.26 $\pm$ 0.09 & 494.60 $\pm$ 0.54 & N/A & N/A & 1.08 $\pm$ 0.001 & 5.94 $\pm$ 0.006 \\
		With polarisation & 6.72 $\pm$ 0.04 & 475.30 $\pm$ 0.67 & 0.50 $\pm$ 0.004 & 58.42 $\pm$ 0.07 & 1.21 $\pm$ 0.002 & 6.73 $\pm$ 0.01 \\
		With polarisation (low freq.) & 5.75 $\pm$ 0.03 & 489.62 $\pm$ 1.03 & 5.37 $\pm$ 0.08 & 74.52 $\pm$ 0.17 & 1.43 $\pm$ 0.004 & 8.02 $\pm$ 0.02 \\
		With polarisation (high freq.) & 2.71 $\pm$ 0.02 & 588.61 $\pm$ 1.17 & 4.06 $\pm$ 0.09 & 106.98 $\pm$ 0.39 & 1.13 $\pm$ 0.002 & 6.22 $\pm$ 0.01 \\
		With polarisation (RFI case 1) & 6.64 $\pm$ 0.04 & 476.39 $\pm$ 0.67 & 0.42 $\pm$ 0.004 & 55.66 $\pm$ 0.11 & 1.33 $\pm$ 0.003 & 7.46 $\pm$ 0.02 \\
		With polarisation (RFI case 2) & 6.76 $\pm$ 0.04 & 482.21 $\pm$ 0.71 & 0.93 $\pm$ 0.01 & 79.70 $\pm$ 0.14 & 1.38 $\pm$ 0.003 & 7.64 $\pm$ 0.02 \\
		With polarisation (RFI case 3) & 6.47 $\pm$ 0.03 & 473.49 $\pm$ 0.67 & 0.66 $\pm$ 0.006 & 62.33 $\pm$ 0.09 & 1.37 $\pm$ 0.003 & 7.59 $\pm$ 0.02 \\
		With polarisation (+lognormal) & 6.58 $\pm$ 0.03 & 475.11 $\pm$ 0.64 & 0.44 $\pm$ 0.004 & 57.86 $\pm$ 0.07 & 2.38 $\pm$ 0.003 & 5.28 $\pm$ 0.006 \\
		\hline
		Prior & $\mathcal{U}(1000,100000000)$ & $\mathcal{U}(200,10000)$ & $\mathcal{U}(0.0001,10)$ & $\mathcal{U}(15,500)$ & $\mathcal{U}(0.000001,0.5)$ & $\mathcal{U}(0.01,15)$ \\
		\hline
	\end{tabular}
    \caption{Median and 1$\sigma$ values of our model hyperparameter posterior distributions for different cases, obtained using nested sampling.}
    \label{table_params}
\end{table*}

We tested the noise scenarios outlined in \secref{gprnoise} and find that the \textit{indistinguishable noise} case performs best. In our data, the noise is so small that the exponential covariance for the \hi signal is probably working well to describe both the \hi signal and the noise. For real data, especially if the noise is larger or more correlated in frequency, this might not be the case.

Our kernel model for this no polarisation case can be summarised as: Exponential (\hi signal kernel, also describing indistinguishable noise) + RBF (smooth foreground kernel) + indistinguishable noise assumed throughout. Our hyperparameters for this model are: $\sigma^2_{\rm smooth}$ is the RBF kernel variance (describing the amplitude of our smooth foregrounds), $\ell_{\rm smooth}$ is the RBF kernel lengthscale (describing the frequency correlation of our smooth foregrounds), $\sigma^2_{21}$ is the exponential kernel variance (describing the amplitude of our \hi signal and indistinguishable noise), and $\ell_{21}$ is the exponential kernel lengthscale (describing the frequency correlation of our \hi signal and indistinguishable noise).

We run nested sampling for this model using \texttt{pymultinest}, obtaining a better estimate for our evidence as well as our hyperparameter posteriors. We find that our evidence obtained with gradient descent (using \texttt{GPy}) and with nested sampling are consistent, as are the best estimates for the hyperparameters (see \autoref{fig:posterior_nopol}). This level of consistency is sufficient for our simulations, where the signal is idealised and the kernel model is simple. For real data the kernel model and hyperparameter posterior distributions will be more complex, so nested sampling should be used instead of gradient descent.

We plot the posterior distribution of our hyperparameters (obtained from nested sampling) in \autoref{fig:posterior_nopol}, and the median and 1$\sigma$ of the distributions can be found in \autoref{table_params}. We also quote the uniform priors imposed in each hyperparameter in \autoref{table_params}. We find that priors are required (especially on lengthscales) to successfully separate the \hi and foreground signal, otherwise the \hi kernel will try to fit to the foregrounds. We tested different prior ranges and found that making it narrower did not improve results, however this might not be the case for real data.

From \autoref{fig:posterior_nopol} we see that the hyperparameters of each kernel (e.g. $\sigma^2_{21}$ and $\ell_{21}$) are correlated with each other but not with those of other kernels, as expected since the signals are different. The dotted lines show the best-fitting hyperparameter estimates from \texttt{GPy}'s gradient descent optimisation, and they agree well with the peaks of the posterior distributions obtained with NS.

\begin{figure*}
	\centering
	\includegraphics[width=2\columnwidth]{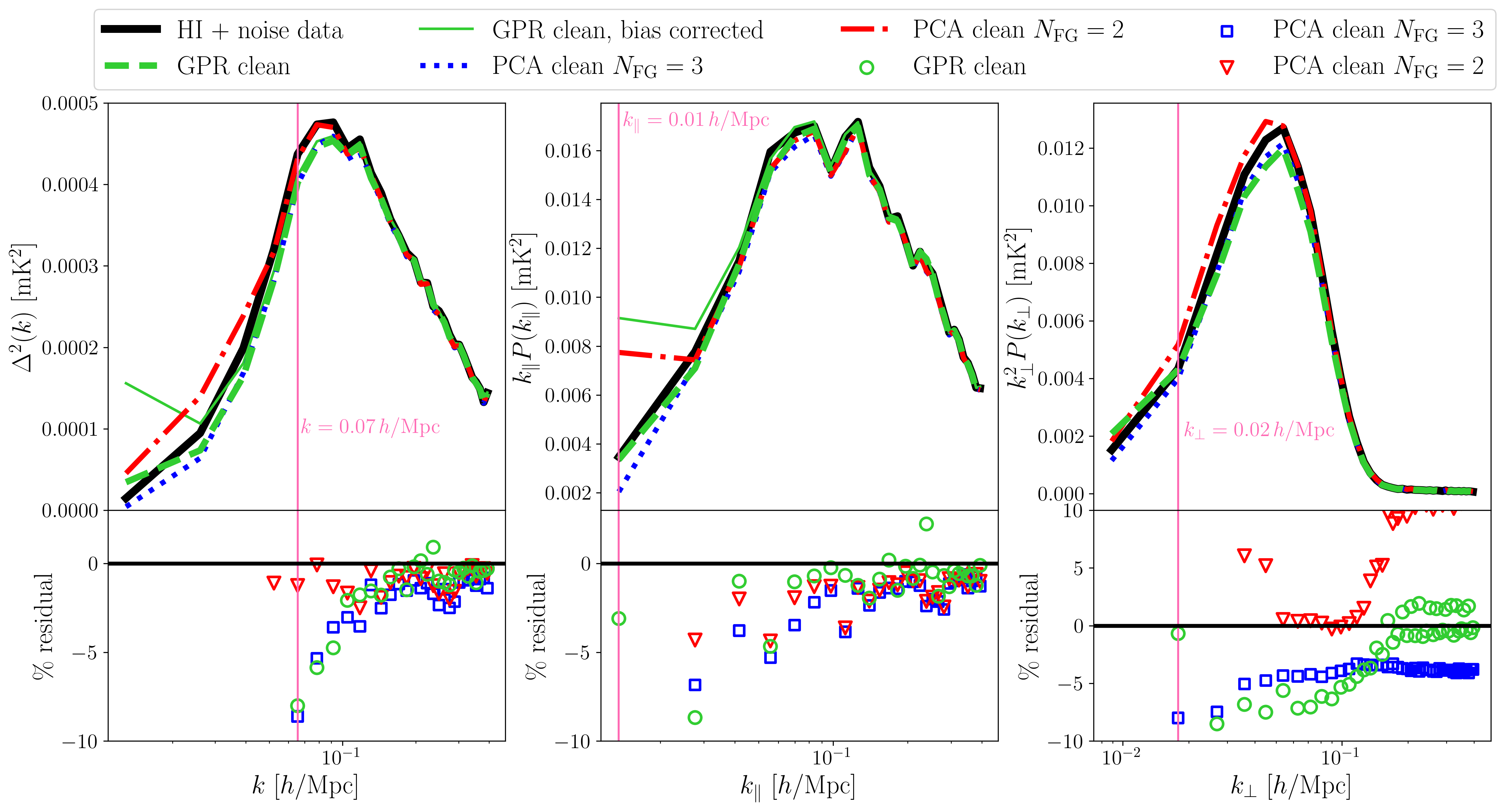}
    \caption{Power spectra results for the no polarisation case. \textit{Top}: True \hi + noise signal (black solid line), PCA foreground cleaned residuals (blue dotted line for $N_{\rm FG} = 3$ and red dash-dotted line for $N_{\rm FG} = 2$), and GPR foreground cleaned residuals without bias correction (green dashed line) and with (thin green solid line). \textit{Bottom}: Percentage residual difference between the foreground removed residual power spectra and the true \hi power spectra. \textit{Left}: Spherically averaged power spectrum. \textit{Centre}: Radial power spectrum. \textit{Right}: Transverse power spectrum. \href{https://github.com/paulassoares/gpr4im/blob/main/Jupyter\%20Notebooks/Reproducible\%20paper\%20plots/No\%20polarisation.ipynb}{\faicon{github-alt}}}
    \label{fig:pk_nopol}
\end{figure*}

We use our best-fitting, optimised kernel model to perform foreground removal with GPR and compare this with PCA. We show two cases for PCA: one with $N_{\rm FG} = 3$, and the other with $N_{\rm FG} = 2$. We plot in \autoref{fig:pk_nopol} the power spectra of these residuals as well as the true residual (\hi and noise) power spectrum. We choose to show the spherically averaged, radial, and transverse power spectra to compare how the foreground removal works in different directions. On the bottom panel we show the percentage difference between the foreground removed power spectra and the true residual power spectra, which we want to be as close to zero as possible.

For the spherically averaged power spectrum, the residuals in \autoref{fig:pk_nopol} show that GPR performs well on small scales, and increasingly worse on large scales. It yields residuals within 10\% of the truth down to a $k$-bin of $k = 0.07\,h/\text{Mpc}$, which we have marked with a vertical pink line. This is larger than the radial and transverse limits ($k_\parallel = 0.01\,h/\text{Mpc}$, $k_\perp = 0.02\,h/\text{Mpc}$) combined, meaning that the individual radial and transverse power spectra can access larger scales than the spherically averaged power spectrum. PCA ($N_{\rm FG} = 3$) also diverges at this $k$-bin, but PCA ($N_{\rm FG} = 2$) recovers larger scales. On small scales, GPR performs better than PCA ($N_{\rm FG} = 3$). While PCA ($N_{\rm FG} = 2$) can access larger scales, it overestimates the power spectrum, which can boost errors.

For the radial power spectrum in \autoref{fig:pk_nopol}, GPR performs remarkably well, able to recover the largest scales down to a $k_\parallel$-bin of $k_\parallel = 0.01\,h/\text{Mpc}$ within 10\% residual. Both PCA cases fail beyond this $k_\parallel$-bin. GPR also recovers small scales better than both PCA cases, making it a superior method for the radial power spectrum recovery.

The thin green line in \autoref{fig:pk_nopol} shows the bias corrected GPR recovery. Including it leads to an over-estimation of the spherically averaged and radial power spectrum at large scales, and no obvious difference on small scales. The transverse power spectrum does not have a bias correction, as discussed in \secref{sec:bias}.

In the transverse power spectrum case in \autoref{fig:pk_nopol}, all methods perform worse than in the radial and spherically averaged power spectrum cases. While GPR can still recover the truth within 10\% residual down to a $k_\perp$-bin of $k_\perp = 0.02\,h/\text{Mpc}$, the residuals tend to lie further away from zero than in the other power spectra cases. GPR again recovers small scales very well, better than PCA. However, since the signal is almost zero here due to the telescope beam, this is not very useful. We attribute the oscillatory nature of the true transverse power spectrum on small scales to the fact that the power here is near zero, so residual differences are floating point precision. Overall, the recovery of the transverse power spectrum is worse than of the radial power spectrum. This makes sense, since GPR mainly takes into account frequency (radial) information.

\subsection{With polarisation}

When including polarised foregrounds, we still find that the exponential kernel is the best fit for the \hi signal, and the RBF kernel is the best for the smooth foregrounds. For the polarised signal, we find that the RBF kernel is also the best fit, though with smaller variance and lengthscale than in the smooth foreground case. We again find that the indistinguishable noise treatment gives the best results.

\begin{figure}
	\centering
	\includegraphics[width=\columnwidth]{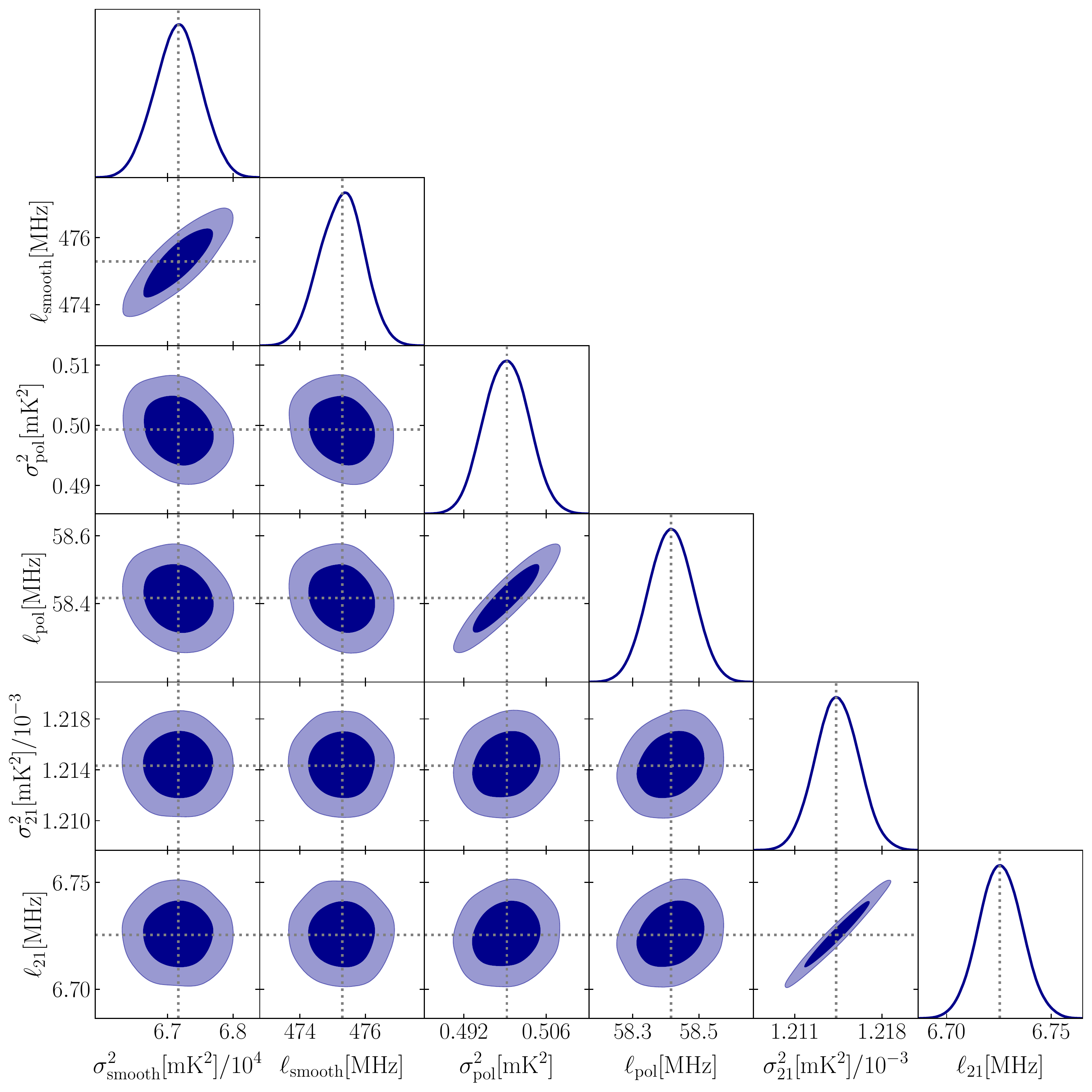}
    \caption{Posterior distribution of our model hyperparameters, obtained using nested sampling in the with polarisation case. Dotted grey lines show the best-fitting hyperparameter results obtained with gradient descent. \href{https://github.com/paulassoares/gpr4im/blob/main/Jupyter\%20Notebooks/Reproducible\%20paper\%20plots/With\%20polarisation.ipynb}{\faicon{github-alt}}}
    \label{fig:posterior_wpol}
\end{figure}

Our kernel model in the presence of polarisation is: Exponential (\hi signal kernel, also describing indistinguishable noise) + RBF (smooth foreground kernel) + RBF (polarised foreground kernel) + indistinguishable noise assumed throughout. We run nested sampling for this model, and plot the hyperparameter posterior distributions in \autoref{fig:posterior_wpol}. As in the no polarisation case, the hyperparameter of each kernel is correlated with itself. The exponential kernel hyperparameters are the most correlated, and the RBF kernel describing the smooth foregrounds has the least correlated hyperparameters. However, the polarised foreground hyperparameters also seem to be negatively correlated with the smooth foreground hyperparameters, and positively correlated with the \hi signal + noise parameters. This indicates that the polarised foreground signal might not be entirely described by the polarised foreground kernel, and some polarised signal is being described by the smooth foreground and \hi + noise kernels. The best-fitting hyperparameter estimates from \texttt{GPy}'s gradient descent again agree well with the posterior distribution obtained with NS. 

We record the median and 1$\sigma$ error estimates of our hyperparameters in \autoref{table_params}. The variance parameter for the smooth foregrounds ($\sigma^2_{\rm smooth}$) is a factor of $\sim$2 smaller than in the no polarisation case, possibly due to difficulty in separating the smooth and polarised foregrounds. The estimated evidence (LML) from nested sampling, describing how well our kernel model describes our data, is smaller when including the polarised foregrounds. For the no polarisation case, the evidence is 1\% larger than when including polarisation leakage.

\begin{figure*}
	\centering
	\includegraphics[width=2\columnwidth]{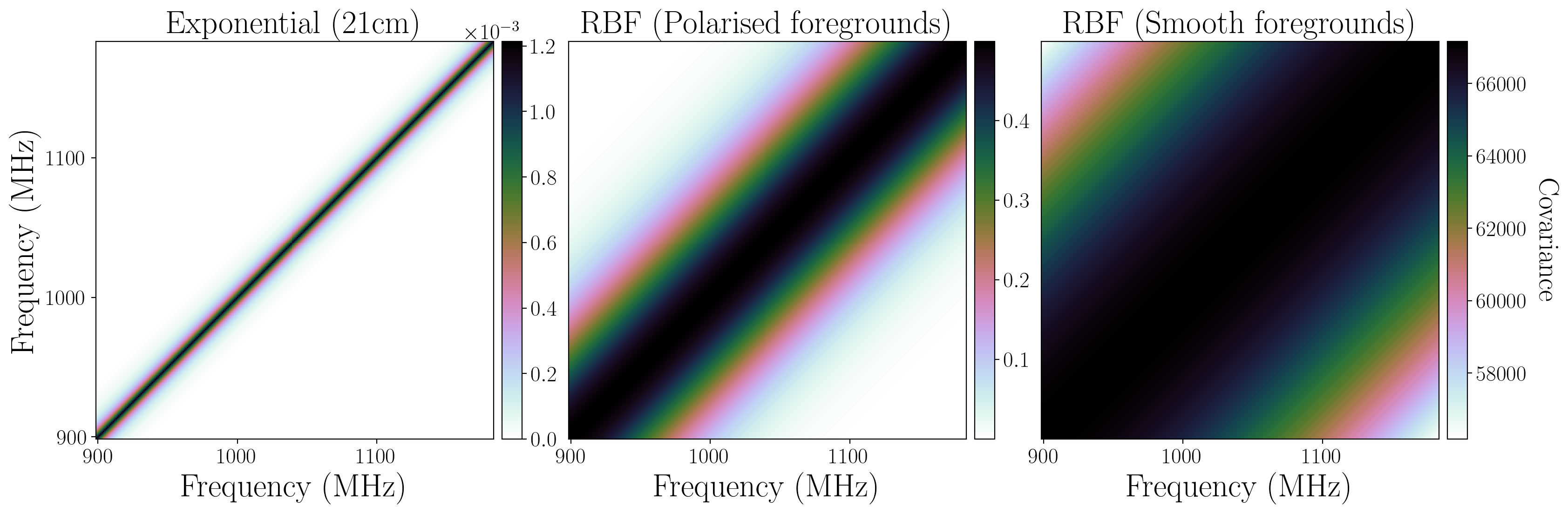}
    \caption{Frequency covariance generated using our best-fit kernel functions, for each signal component. These are generated for the frequency range of our data, using our best estimate for the kernel hyperparameters from nested sampling. \href{https://github.com/paulassoares/gpr4im/blob/main/Jupyter\%20Notebooks/Reproducible\%20paper\%20plots/Assorted\%20plots.ipynb}{\faicon{github-alt}}}
    \label{fig:cov_bestfit}
\end{figure*}

We plot the best-fitting kernel function for each component in \autoref{fig:cov_bestfit}. The \hi exponential kernel is much less correlated in frequency than either foreground kernel, and the smooth foreground RBF kernel is the most correlated in frequency. From the amplitude of the colour bar, the \hi exponential kernel has the smallest variance, while the smooth foreground RBF kernel has the largest variance (and consequently signal amplitude), as expected.

\begin{figure*}
	\centering
	\includegraphics[width=2\columnwidth]{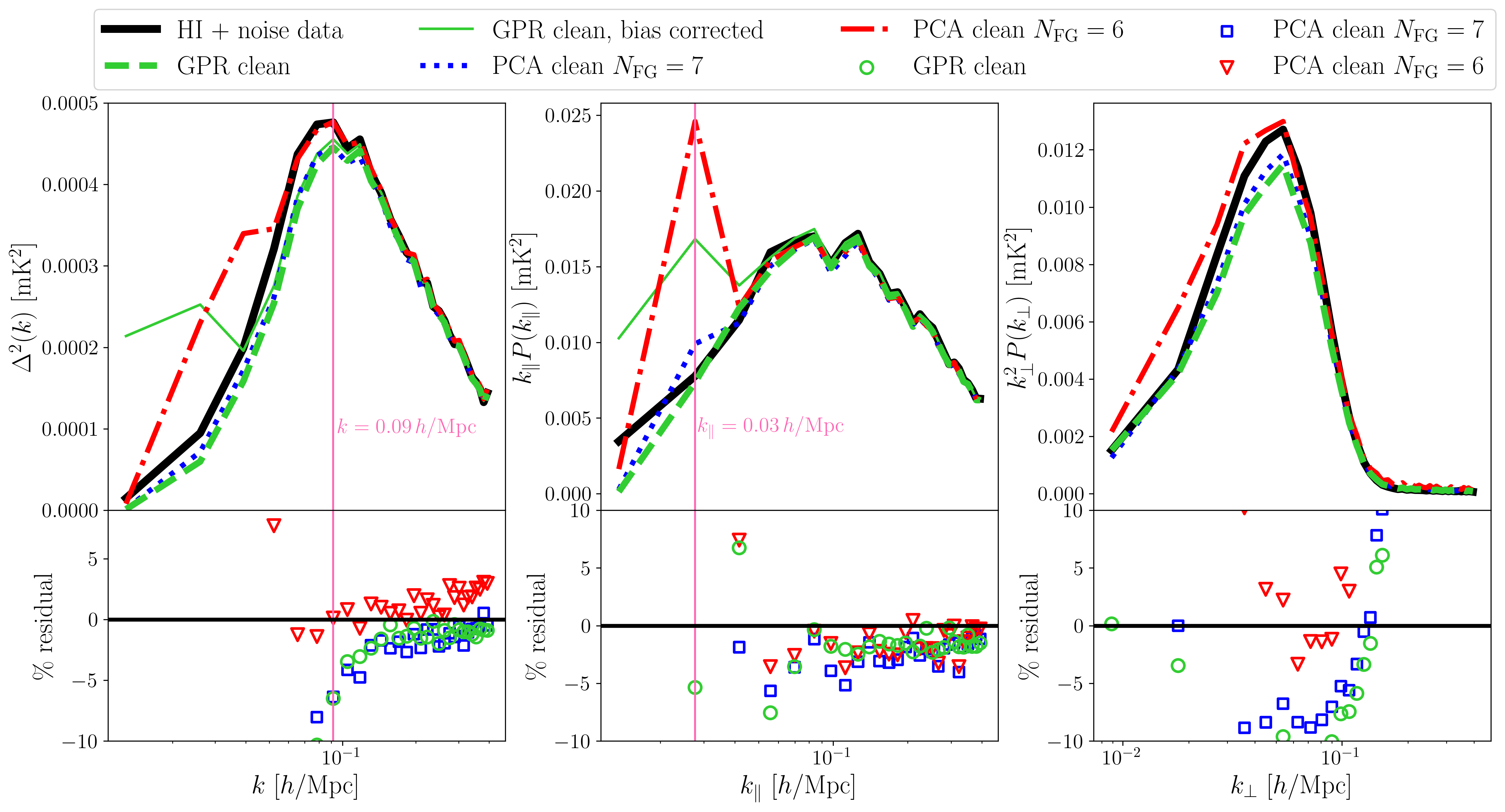}
    \caption{
    Power spectra results for the case including polarisation. \textit{Top}: True \hi + noise signal (black solid line), PCA foreground cleaned residuals (blue dotted line for $N_{\rm FG} = 7$ and red dash-dotted line for $N_{\rm FG} = 6$), and GPR foreground cleaned residuals without bias correction (green dashed line) and with (thin green solid line). \textit{Bottom}: Percentage residual difference between the foreground removed residual power spectra and the true \hi power spectra. \textit{Left}: Spherically averaged power spectrum. \textit{Centre}: Radial power spectrum. \textit{Right}: Transverse power spectrum. \href{https://github.com/paulassoares/gpr4im/blob/main/Jupyter\%20Notebooks/Reproducible\%20paper\%20plots/With\%20polarisation.ipynb}{\faicon{github-alt}}}
    \label{fig:pk_wpol}
\end{figure*}

We perform foreground removal with this best-fit GPR model, and plot our residual power spectra in \autoref{fig:pk_wpol}. We also show the true residual power spectrum, as well as the residual obtained with PCA (now looking at the cases with $N_{\rm FG} = 7$ and $N_{\rm FG} = 6$, since the polarisation leakage foregrounds require a larger $N_{\rm FG}$ to be properly removed). The bottom panel shows the percentage residual difference between the foreground cleaned and true residuals.

For the spherically averaged power spectrum, \autoref{fig:pk_wpol} shows that GPR performs better than both PCA cases on small cases, since the \% residuals are closer to zero and it does not overestimate the power spectrum. The residuals get gradually worse for larger scales, until they diverge beyond 10\% below a $k$-bin of $k = 0.09\,h/\text{Mpc}$ . This a larger limit than in the no polarisation case, indicating that the polarised foregrounds make the cleaning more difficult. From the residuals, PCA ($N_{\rm FG} = 7$) is able to access larger scales than GPR. PCA ($N_{\rm FG} = 6$) overestimates the residual power spectrum on both small and large scales.

For the radial power spectrum, GPR is once again good at recovering the truth, with residuals below 10\% down to $k_\parallel = 0.03\,h/\text{Mpc}$ (a larger cut-off than in the no polarisation case). The PCA residuals diverge after this point, meaning GPR can access larger scales. For the PCA $N_{\rm FG} = 6$ case, influence from residual foregrounds is strong and they are particularly dominant in the second smallest $k_\parallel$ mode. The transverse power spectrum results are dramatically worse than in the no polarisation case. Neither PCA or GPR recover the transverse power spectrum of the true signal well. This is probably due to how the foreground removal works in both cases: it mainly takes into account frequency (radial) information, so it is worse at recovering the true residual transverse power spectrum.

Once again, we see in \autoref{fig:pk_wpol} that including the bias correction leads to an overestimation of the spherically averaged and radial power spectra on large scales, and no clear difference on small scales.

\begin{figure}
	\centering
	\includegraphics[width=\columnwidth]{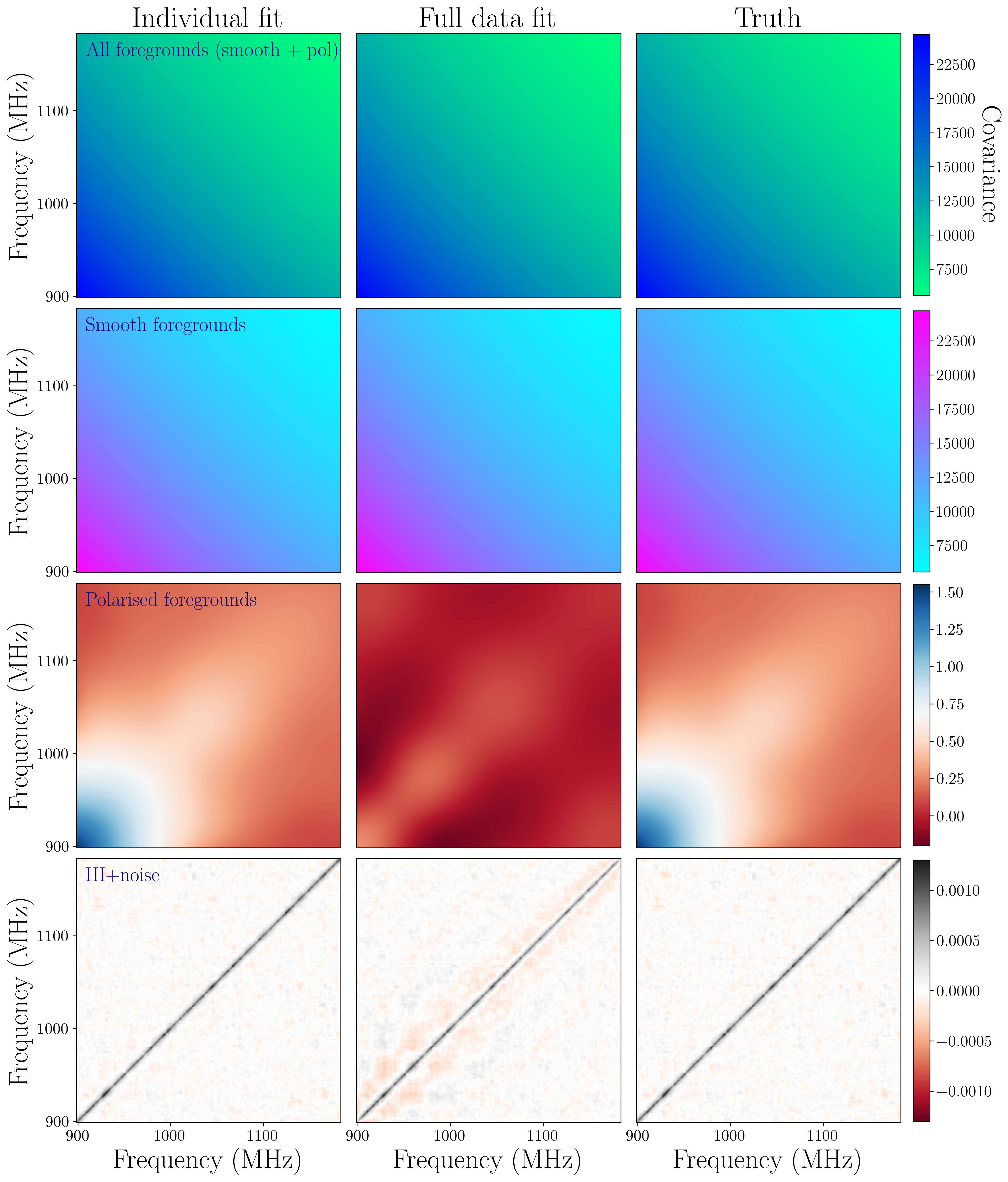}
    \caption{Frequency covariance of our signals, as predicted by GPR and compared to the truth. We look at all foregrounds together (\textit{top row}), the smooth foregrounds only (\textit{second row}), the polarised foregrounds only (\textit{third row}), and the \hi+noise only (\textit{bottom row}). \textit{Left}: GPR prediction of what the signal frequency covariance looks like, given our chosen kernels and based on the relevant signal data only. \textit{Centre}: GPR prediction of what the signal frequency covariance looks like, given our chosen kernels and based on our full data. \textit{Right}: The true frequency covariance of our data. \href{https://github.com/paulassoares/gpr4im/blob/main/Jupyter\%20Notebooks/Reproducible\%20paper\%20plots/With\%20polarisation.ipynb}{\faicon{github-alt}}}
    \label{fig:fg_covs}
\end{figure}

To better understand how well our best-fitting kernel functions are describing the signals in our data, we compare the true frequency covariance of our signals to what GPR predicts. In \autoref{fig:fg_covs}, we look at two different cases of what GPR recovers, described below.

First, we take the RBF kernel for the smooth foregrounds and the RBF kernel for the polarised foregrounds, and use our full data to predict what the foreground signal looks like, i.e. we are estimating $\text{E}[\textbf{f}_{\rm fg}]$. This is what we've already done previously to predict the foreground signal and remove it from our data, but we are focusing now on what this prediction looks like. We do this also for the smooth foreground RBF kernel, to predict what the smooth foregrounds look like (i.e. $\text{E}[\textbf{f}_{\rm smooth}]$). We do the same for the polarised foregrounds RBF kernel, obtaining an estimate for what our polarised foregrounds look (i.e. $\text{E}[\textbf{f}_{\rm pol}]$). Finally, we also do this for the \hi signal and indistinguishable noise kernel. These are denoted as a ``full data fit'', since we are using the kernels and the full data to predict the signals. This ``full data fit'' is what we've already done to remove foregrounds using GPR, but here we are focusing on what the signal predictions looks like, and how it compares to the true signal. We are also considering a new case of only fitting the smooth foreground kernel, and also only fitting the polarised foreground kernel, to try to predict these signals in isolation.

We also consider an ``individual fit'', where instead of fitting our kernels to our full data, we fit them to the appropriate data \textit{only}. For the RBF + RBF full foreground kernel, we fit this to the smooth + polarised foreground data. For the RBF smooth foreground kernel, we fit it only to the smooth foreground data, and similarly for the polarised foreground case. For the Exponential kernel, we fit it only to the \hi and noise data. This ``individual fit'' is a new analysis, which we are only doing for understanding GPR better, and it does not play a part in our foreground removal pipeline. The only difference from the ``full data fit'' is that instead of calculating the signal estimates using the full data, we only use the relevant signal data (i.e., we are predicting the foregrounds given the foreground kernels and foreground data, without the need to separate it from the \hi signal or noise).

We expect that the ``individual fit'' will give predictions that are closer to the truth than the ``full data fit'', since in this case we are not trying to separate our signals, but are instead fitting to the isolated signals. This is indeed what we find in \autoref{fig:fg_covs}, where we plot the frequency covariance of our predictions as well as the real covariance. The left panel (``individual fit'') looks similar to the right panel (truth), while the middle panel (``full data fit'') shows some disagreement. For the all foregrounds and smooth foregrounds case, the difference between these cases is not too stark. However, for the polarised foregrounds and \hi+noise signal it is. We can see that the ``full data fit'' for the polarised foregrounds is not great, since it does not match the truth so well. The fact that the ``full data fit'' is bad, but the ``individual fit'' is good means that the issue is not with our best-fitting foreground kernels. The issue arises when we try to pick out the polarised signal in the presence of other signals - this makes it easier for the signals to be confused with each other, and leads to more uncertainty. Our foreground kernels are good fits for the signals individually, but it is difficult to isolate the polarised foregrounds in the presence of other signals.

The difficulty in recovering the \hi+noise signal in the ``full data fit'' case shows that foreground removal in the presence of polarised foregrounds is more difficult. In the ``individual fit'' case, the \hi+noise covariance recovery matches the truth well, but when considering all signals together in the ``full data fit'', this is no longer true. This suggests that the \hi+noise signal is being confused with other signals, much like the polarised foregrounds. This confusion and inability to separate signals makes the final foreground prediction more uncertain, and ultimately hinders the foreground removal process.

\subsection{Bandwidth and redshift dependence}

We wanted to test how our results would differ if we cut our data's bandwidth in half, i.e. only considered half of our frequency range. Since GPR uses frequency information, it might be more difficult for it to fit the data well if there is less frequency data to learn from. Alternatively, as suggested by \cite{Hothi_2020}, it might be better to consider narrower frequency ranges in the case where the signal's lengthscale is changing with frequency, that way it avoids being ``averaged out'' as much.

We test this in the case of including polarised foregrounds, and again find the best model to be: Exponential (\hi signal kernel, also describing indistinguishable noise) + RBF (smooth foreground kernel) + RBF (polarised foreground kernel) + indistinguishable noise assumed throughout. The only difference now is that when we predict the foreground signal from our data using this model, we only consider half of the frequency range. When we split our data in half along frequency, we also compare both halves, to see if the redshift/frequency of observation makes a difference. 

\begin{figure*}
	\centering
	\includegraphics[width=2\columnwidth]{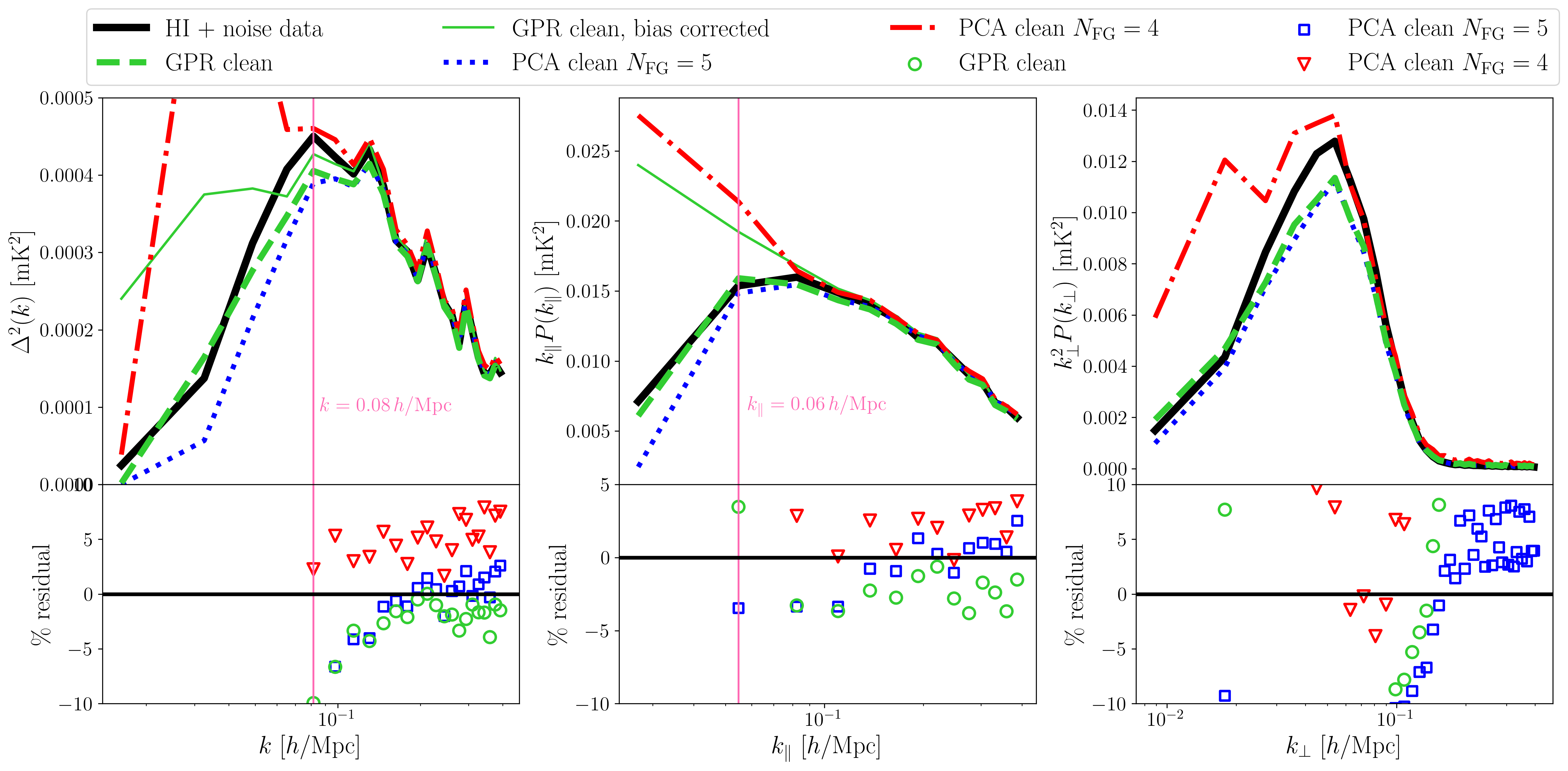}
    \caption{Power spectra results for the low frequency half-bandwidth case including polarisation. \textit{Top}: True \hi + noise signal (black solid line), PCA foreground cleaned residuals (blue dotted line for $N_{\rm FG} = 5$ and red dash-dotted line for $N_{\rm FG} = 4$), and GPR foreground cleaned residuals without bias correction (green dashed line) and with (thin green solid line). \textit{Bottom}: Percentage residual difference between the foreground removed residual power spectra and the true \hi power spectra. \textit{Left}: Spherically averaged power spectrum. \textit{Centre}: Radial power spectrum. \textit{Right}: Transverse power spectrum. \href{https://github.com/paulassoares/gpr4im/blob/main/Jupyter\%20Notebooks/Reproducible\%20paper\%20plots/With\%20polarisation\%20(bandwidth\%20and\%20redshift\%20dependence).ipynb}{\faicon{github-alt}}}
    \label{fig:pk_wpol_lofreq}
\end{figure*}

\begin{figure*}
	\centering
	\includegraphics[width=2\columnwidth]{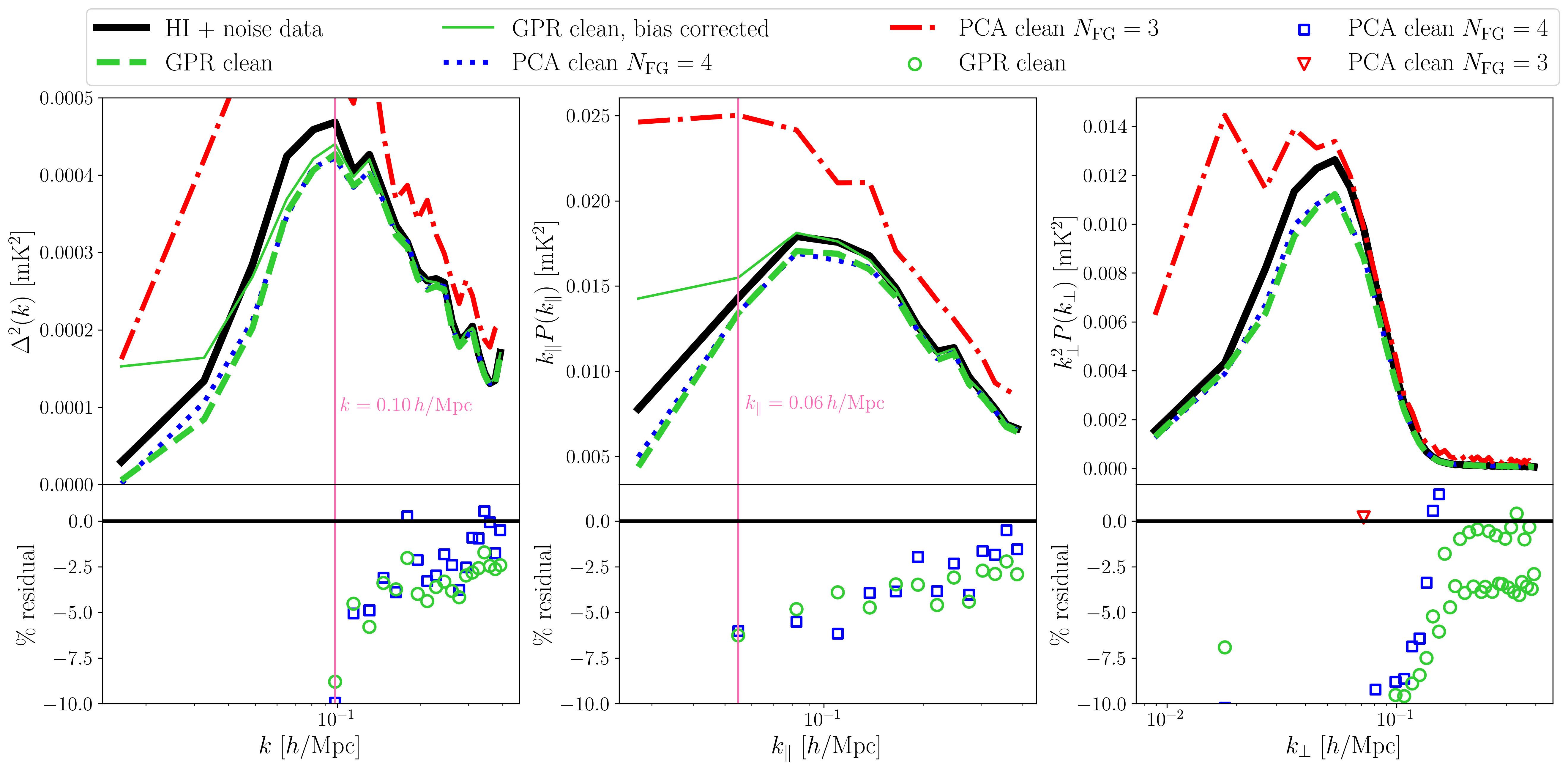}
    \caption{Power spectra results for the high frequency half-bandwidth case including polarisation. \textit{Top}: True \hi + noise signal (black solid line), PCA foreground cleaned residuals (blue dotted line for $N_{\rm FG} = 4$ and red dash-dotted line for $N_{\rm FG} = 3$), and GPR foreground cleaned residuals without bias correction (green dashed line) and with (thin green solid line). \textit{Bottom}: Percentage residual difference between the foreground removed residual power spectra and the true \hi power spectra. \textit{Left}: Spherically averaged power spectrum. \textit{Centre}: Radial power spectrum. \textit{Right}: Transverse power spectrum. \href{https://github.com/paulassoares/gpr4im/blob/main/Jupyter\%20Notebooks/Reproducible\%20paper\%20plots/With\%20polarisation\%20(bandwidth\%20and\%20redshift\%20dependence).ipynb}{\faicon{github-alt}}}
    \label{fig:pk_wpol_hifreq}
\end{figure*}

We again use nested sampling to find the best-fitting hyperparameters of our model, and note down the median and 1$\sigma$ of the distributions in \autoref{table_params}. We then perform foreground removal using GPR, and plot the residual power spectra results for the low frequency, high redshift case in \autoref{fig:pk_wpol_lofreq}, and for the high frequency, low redshift case in \autoref{fig:pk_wpol_hifreq}.

\begin{figure*}
	\centering
	\includegraphics[width=2\columnwidth]{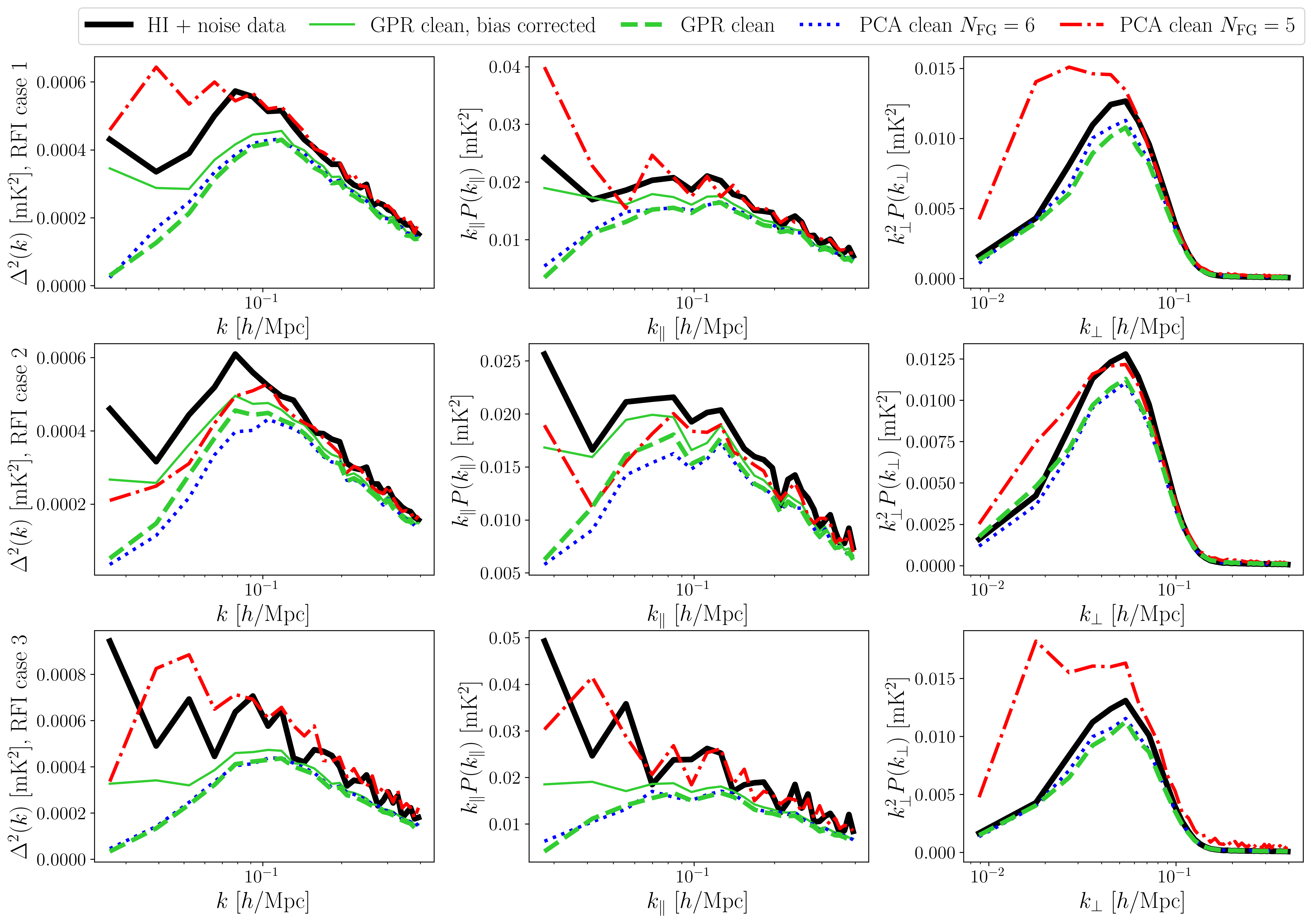}
    \caption{Power spectra results for the RFI-like missing channels cases including polarisation. For all cases, 40\% of the frequency channels are missing, but each row represents a different distribution of missing channels in the data. We show the true \hi + noise signal (black solid line), PCA foreground cleaned residuals (blue dotted line for $N_{\rm FG} = 6$ and red dash-dotted line for $N_{\rm FG} = 5$), and GPR foreground cleaned residuals without bias correction (green dashed line) and with (thin green solid line). \textit{Left}: Spherically averaged power spectrum. \textit{Centre}: Radial power spectrum. \textit{Right}: Transverse power spectrum. \href{https://github.com/paulassoares/gpr4im/blob/main/Jupyter\%20Notebooks/Reproducible\%20paper\%20plots/With\%20polarisation\%20(RFI\%20gaps).ipynb}{\faicon{github-alt}}}
    \label{fig:pks_rfi}
\end{figure*}

For the low frequency case, we show the $N_{\rm FG} = 5$ and $N_{\rm FG} = 4$ cases for PCA. For the spherical power spectra of \autoref{fig:pk_wpol_lofreq}, GPR tends to lie closer to the true residual power spectrum line, indicating it is performing better than PCA. In small scales, both PCA cases overestimate the true residual power spectrum, while GPR slightly underestimates. GPR recovers the power spectrum within 10\% residuals down to a $k$-bin of $k = 0.08\,h/\text{Mpc}$, smaller than the full bandwidth case, indicating that it can now access larger scales. GPR also recovers the radial power spectrum nicely, but not the transverse power spectrum, as before.

For the high frequency case, we show the PCA results for $N_{\rm FG} = 4$ and $N_{\rm FG} = 3$. GPR does not outperform PCA in this case, but agrees nicely with the PCA $N_{\rm FG} = 4$ recovery throughout. It now recovers the spherically averaged power spectrum within 10\% residual down to $k = 0.1\,h/\text{Mpc}$, which is larger than the low frequency case and full bandwidth case. GPR performs worse here than in the low frequency case. As seen in \autoref{fig:data_samples}, the amplitude of our foregrounds is higher in the lower frequency range than in the higher frequency range, and this could make a difference in our GPR foreground removal results.

In both cases, including the bias correction results in an overestimation of the spherically averaged and radial power spectra on large scales, but some discernible difference on medium to small scales that makes it closer to the truth. It is interesting that the low frequency half experiences a more extreme bias correction.

\cite{Hothi_2020} found that GPR performs worse in larger bandwidths. Here this is not so clear - the low frequency case can access larger scales than the full bandwidth case, but the high frequency case performs worse than the full bandwidth one. \cite{Hothi_2020} considers interferometric EoR simulations of \hi signal, which has different systematics to our single-dish, low redshift IM simulations. This, and the fact that their noise level is much larger than ours, could explain why we don't see the same effects.

\subsection{RFI effects}

It is common in \hi IM experiments to have to remove whole frequency channels due to them being contaminated with RFI. We test how GPR performs when frequency channels are missing, as compared to PCA, in the case of including polarised foregrounds. To do this, we use the method described in \cite{Carucci_2020}, where three different cases are considered in the context of the sparsity-based foreground removal algorithm generalized morphological component analysis (GMCA; \citealt{Chapman_2012}): 

\begin{itemize}[leftmargin=*]
    \item \textbf{Case 1}: The case of removing 40\% of the channels in one chunk in the middle of the frequency range, such that we have in order: 30\% good channels, 40\% bad (missing) channels, and 30\% good channels;
    \item \textbf{Case 2}: The case of removing 40\% of the channels in one chunk at the beginning of the frequency range, such that we have in order: 10\% good channels, 40\% bad (missing) channels, and 50\% good channels;
    \item \textbf{Case 3}: The case of removing 40\% of the frequency channels, in a few different chunks throughout the frequency range, such that we have in order: 20\% good channels, 30\% bad (missing) channels, 20\% good channels, 10\% bad (missing) channels, and 20\% good channels.
\end{itemize}

Once again we used the kernel model: Exponential (\hi signal kernel, also describing indistinguishable noise) + RBF (smooth foreground kernel) + RBF (polarised foreground kernel) + indistinguishable noise assumed throughout, and ran nested sampling to obtain the best-fitting kernel hyperparameters (see \autoref{table_params} for the best-fitting median and 1$\sigma$ values). The only difference from our full data case is that here we are ignoring certain channel when running GPR and performing foreground removal. We are not setting them to zero, but instead removing the channels entirely from the data, thus reducing the number of frequency channels $N_\nu$.

We plot our resulting power spectra for each of these different cases in \autoref{fig:pks_rfi}. We show the cases of $N_{\rm FG} = 6$ and $N_{\rm FG} = 5$ for PCA. For the spherically averaged power spectra on the top row, GPR performs worse than it did in the no missing channels case, and either performs the same as or worse than the best PCA case. The same is true for the radial power spectrum in the middle row. The bias correction in this case does improve the spherically averaged and radial power spectra results, without leading to an overestimation on large scales as seen previously.

The foreground removal pipeline for GPR depends on first modelling the foreground signal as a function of frequency, so having gaps in the input data significantly decreases the continuity and quality of information GPR has to learn from. This is turn leads to worse foreground removal than in the no missing channels case, and worse foreground removal than PCA in many cases, since PCA does not have this strong dependency on modelling the foregrounds as a function of frequency.

Interestingly, the transverse power spectrum does not see a dramatic difference, and looks similar to the recovered transverse power spectra in the no missing channels case. This is not surprising, since there is no gap of information in the spatial direction, only in the frequency direction. 

While both GPR and PCA are negatively impacted by missing frequency channels, \cite{Carucci_2020} found that GMCA is still able to recover the true radial power spectrum in the presence of these complications. \cite{Carucci_2020} also tested this for the Independent Component Analysis algorithm FastICA and found that, much like PCA and GPR, it is compromised by missing frequency channels.

\cite{Offringa_2019} studied how missing frequency channels due to RFI affect GPR as a foreground removal technique in the context of EoR studies. They use an interpolation scheme to correct for the flagged RFI channels, which differs from our method of simply excluding the channels, and found that RFI flagging leads to excess power in the final foreground removed power spectrum. We leave investigation of how interpolation methods for missing frequency channels would affect our results to future work.

\subsection{Foreground transfer function}

Implementing a foreground removal transfer function is a common way of mitigating foreground removal effects that might be present in the residual power spectrum \citep{Switzer_2015}. It is usually done by injecting \hi IM mocks into the data, performing the foreground removal, and cross-correlating the result with the original mock, in order to obtain an estimate for how the signal has been biased \citep{Masui:2012zc,Switzer_2013,Switzer_2015,Anderson_2018, cunnington202021cm, wolz2021hi}. This works well for blind foreground removal methods such as PCA, since the mocks should be as uncorrelated in frequency as the true \hi signal. Therefore, both the mocks and the true \hi signal will lose similar modes (the ones most degenerate with the foregrounds) in the foreground clean.

Here we begin to investigate whether this transfer function technique could work for GPR. With GPR, we are modelling each component of our signal with a different kernel, and so we wanted to check if we would need to adapt our kernel model in the presence of mock data. We use a lognormal simulation of cosmological \hi signal as our mock. We do this in the case of including polarised foregrounds.

We generated a lognormal realisation of our cosmological signal (a method first proposed by \cite{Coles:1991}), in the same way as described in e.g. \cite{Cunnington_2020nongauss}. We then smooth it by the same telescope beam as before, and add it to our data. Our data is now composed of the smooth and polarised foregrounds, our original \hi signal, this lognormal \hi realisation, and noise.

We tested whether the model: Exponential (\hi signal kernel, also describing indistinguishable noise) + RBF (smooth foreground kernel) + RBF (polarised foreground kernel) + indistinguishable noise assumed throughout would still work well with this data, or if we need an extra exponential kernel to describe the lognormal realisation. By comparing the evidence in both cases, we find that the original model with only one exponential kernel is preferred, probably due to how difficult it is for GPR to pick out very similar signals, making it better to model them together (this is similar to how it is better to model the \hi signal and noise together with one exponential kernel).

We ran nested sampling using our original model and this new data including a lognormal realisation. The median and 1$\sigma$ distributions of our hyperparameters can be found in \autoref{table_params}. The variance of the exponential kernel is larger than for the no-lognormal case (almost double), which makes sense since we have essentially doubled our \hi signal smplitude. The lengthscale remains constant, because the signals are similar. 

We compared how GPR foreground removal works in this case, to the no-lognormal case. We perform foreground removal on our data with the lognormal signal, using this best-fit model, and find that the residual power spectrum looks extremely similar in shape to the no-lognormal case, only changing in amplitude since it includes the extra lognormal realisation. Adding a lognormal realisation of signal does not make the foreground removal any worse, or overall different. Most importantly, it does not require a new or different kernel function to describe it. This is good, since it means we can use the foreground transfer function technique in the context of GPR. Further work is required to determine whether it is viable to apply both a bias correction \textit{and} a transfer function to GPR residuals. 

\section{Conclusions}\label{sec:conclusion}

The take-home message of this paper is that GPR can be used as a foreground removal technique for low redshift, single-dish \hi intensity mapping. We presented a pipeline for performing GPR in this case, which uses the data to optimise and choose a kernel model. For PCA, we had to fine-tune the $N_{\rm FG}$ parameter for each case considered, and found that even just decreasing the bandwidth requires a different $N_{\rm FG}$ choice. This is a problem for real data, where we don't know the truth and have to make an educated guess for what the best $N_{\rm FG}$ choice is. GPR operates differently - it uses the data to determine the best choice for the kernel model, and then performs the foreground removal. We do not have to make an educated guess with GPR, as the Bayes factor analysis guides our final choice. This poses a significant advantage over blind foreground removal methods that require a fine-tuning parameter such as $N_{\rm FG}$ (e.g. PCA, FastICA, GMCA).

We presented the publicly available and user friendly \texttt{python} package gpr4im\footnote{\href{https://github.com/paulassoares/gpr4im}{github.com/paulassoares/gpr4im}}, which can be used to run GPR as a foreground removal technique in any real space \hi intensity map, either simulated or real. Although we considered here the single-dish \hi IM case, our code could also be used in the context of interferometric data, and higher redshift signal.

Our main findings can be summarised as:

\begin{itemize}[leftmargin=*]
    
    \item GPR may be used as a foreground removal technique in the single-dish, low redshift \hi IM case. Without polarisation leakage in the foregrounds, it recovers the true residual power spectrum within 10\% residual difference down to a $k$-bin of $k = 0.07\,h/\text{Mpc}$. With polarisation leakage, it does so down to $k = 0.09\,h/\text{Mpc}$, since the polarised foregrounds make the removal more difficult;\\
    
    \item In the no polarisation case, GPR recovers the BAO scales better than PCA in the radial power spectrum, but worse in the transverse case. For the spherically averaged power spectrum, GPR recovers the small BAO scales best, but PCA does better on the larger BAO scales. The same is true for the case including polarisation, except for the radial power spectrum, where GPR no longer outperforms PCA.\\
    
    \item GPR outperforms PCA at recovering the true residual power spectrum on small scales, and both fail on large scales. This is no longer true when the bandwidth is halved of frequency channels are missing;\\
    
    \item GPR is very good at recovering the true residual radial power spectrum, more so than PCA in the full bandwidth case. Both GPR and PCA are worse at recovering the transverse power spectrum, due to how they mainly take into account frequency information;\\
    
    \item GPR prefers lower frequencies. When splitting the bandwidth of our data in half, GPR performs much better in the low frequency (high redshift) domain, where the foregrounds are brighter, and recovered the power spectrum within 10\% residual down to a $k = 0.08\,h/\text{Mpc}$. In the high frequency case, this went up to $k = 0.1\,h/\text{Mpc}$;\\
    
    \item In the presence of missing channels due to RFI contamination, both PCA and GPR suffer, and GPR either performs the same as or worse than PCA on all scales. This is clear for the power spectrum and radial power spectrum. The transverse power spectrum is not much affected in any case;\\
    
    \item The \cite{Mertens_2020} bias correction causes an overestimation of the spherically averaged and radial power spectrum on large scales. However, in the case when frequency channels are missing, the bias correction helps us recover the truth better. When halving the bandwidth, the bias correction improves results on medium to small scales.\\
    
    \item Adding a lognormal realisation of \hi cosmological signal does not require an additional kernel, and does not change how GPR performs as a foreground removal technique. As such, it should be possible to apply a foreground removal transfer function with GPR.

\end{itemize}

We hope that these results are useful for implementing GPR as a foreground removal technique in the single-dish, low redshift \hi IM case going forward. Future plans include performing a foreground removal method comparison study like this one on real data, to understand how GPR performs in the context of real \hi IM data.

It would also be interesting to try different power spectrum bias correction techniques on the residual power spectrum, such as the one described in \cite{kern2020gaussian}. It could also be beneficial to try to incorporate the error in the hyperparameter posterior distributions obtained with nested sampling into the residual power spectrum estimation.

Since the smooth foreground signal is actually composed of three different signals (Galactic synchrotron, free-free and point source emission), it would also be interesting to check whether three kernels instead of one are best for describing the smooth foregrounds. 

Finally, we reiterate that our kernel models used in this paper are overly optimistic since our simulations are idealised, and a full nested sampling model selection procedure should be performed when choosing the best-fitting kernels for real data.

\section*{Acknowledgements}

We are grateful to the reviewer for their extremely helpful comments that improved the quality of this manuscript. We thank Ian Hothi, Phil Bull, Florent Mertens, Isabella Carucci, Mario Santos, Jos{\'e} Fonseca for helpful feedback and Andrei Mesinger, Fraser Kennedy and Ioannis Patras for useful discussions. PS is supported by the Science and Technology Facilities Council [grant number ST/P006760/1] through the DISCnet Centre for Doctoral Training. CW's research for this project was supported by a UK Research and Innovation Future Leaders Fellowship, grant number MR/S016066/1. SC is supported by STFC grant ST/S000437/1. AP is a UK Research and Innovation Future Leaders Fellow, grant MR/S016066/1, and also acknowledges support by STFC grant ST/S000437/1. This research utilised Queen Mary’s Apocrita HPC facility, supported by QMUL Research-IT \url{http://doi.org/10.5281/zenodo.438045}. We acknowledge the use of open source software \citep{scipy:2001, Hunter:2007,  mckinney-proc-scipy-2010, numpy:2011,  Lewis1999bs,Lewis2019xzd}. Some of the results in this paper have been derived using the \texttt{healpy} and \texttt{HEALPix} package. We thank New Mexico State University (USA) and Instituto de Astrofisica de Andalucia CSIC (Spain) for hosting the Skies \& Universes site for cosmological simulation products.\\

\section*{Data Availability}

The data and code underlying this article can be found \href{https://github.com/paulassoares/gpr4im}{here}.




\bibliographystyle{mnras}
\bibliography{Bib} 





\appendix

\section{Simulations}\label{SimAppendix}

\subsection{Smooth foregrounds}

Here we describe how we generate the smooth foregrounds in our simulation. The most abundant smooth foregrounds that pose problems in \hi IM experiments are synchrotron emission, free-free emission and point sources.  The synchrotron emission comes from our Galaxy's magnetic field accelerating cosmic-ray electrons, causing them to emit radiation. The free-free emission comes mainly from  ions in our Galaxy causing free electrons to scatter, but some extragalactic contribution to this signal is present. The point sources represent extragalactic objects which emit radio signals, such as Active Galactic Nuclei.

For the Galactic synchrotron and Galactic free-free emission, we use the Planck Legacy Archive\footnote{\href{http://pla.esac.esa.int/pla/}{pla.esac.esa.int/pla}} FFP10 simulations. For more detail on these simulations, please see C21.

\subsubsection{Synchrotron emission}

To generate maps of Galactic synchrotron emission, we take the FFP10 maps at frequencies of 217 and 353 GHz. These maps are derived from the source-subtracted and destriped 408 MHz all-sky map, which is limited by a resolution of 56 arcmin but this $N_{\rm side} = 2048$ version uses Gaussian random field realisations to fill in the higher resolution details \citep{remazeilles2015improved}.

The synchrotron spectral index map can be determined from the 217 and 353 GHz synchrotron maps, which use the 'Model 4' synchrotron spectral index map \citep{MivilleDeschenes}. This has a resolutions of approximately 5 degrees, so we must fill in the higher resolution details. To do this, we use a Gaussian random field, using the synchrotron scaling relation in \cite{Santos_2005}. We then use this spectral index map to interpolate the synchrotron emission at our frequency range.

\subsubsection{Free-free emission}

For simulating the Galactic free-free emission, we once again make use of the FFP10 maps, specifically the 217 GHz free-free simulation which has $N_{\rm side} = 2048$ \citep{MivilleDeschenes}. This map is made using the free-free template from \cite{Dickinson_2003} as well as the ones from WMAP MEM. Taking the free-free amplitude ($a_{\rm free}$) from this map, we then simulate the free-free emission using a power law
\begin{equation}
    T_{\rm free} = a_{\rm free} \left( \frac{\nu}{\nu_0} \right)^{\beta_{\rm free}} \, ,   
\end{equation}
where $\beta_{\rm free}$ is the free-free spectral index, which is constant for all pixels.

\subsubsection{Point sources}

The point sources are generated by fitting a polynomial to radio sources at 1.4 GHz, using the model described in \cite{Battye_2013}. We then use the method described in \cite{Olivari_2017} to scale this signal down to our frequencies. This method uses a power law, where the spectral index is described by a Gaussian distribution with a mean of -2.7 and standard deviation of 0.2. Our source extraction has an upper bound of 100 mJy.

\subsection{Polarisation leakage}

Polarisation leakage arises from the magnetic fields present in our Galaxy's interstellar medium. When light interacts with these fields, Faraday rotation can change its polarisation angle. This happens for our Galactic synchrotron emission, and some of its emission will be leaked from Stokes Q and Stokes U to Stokes I. This leakage will change with frequency because Faraday rotation is frequency dependent \citep{jelic2010, Moore_2013}. Any frequency-dependent variation to the foreground signal will make foreground removal more difficult, specifically for methods that rely on foregrounds being smooth in frequency.

To simulate this effect, we use \texttt{CRIME}\footnote{\href{http://intensitymapping.physics.ox.ac.uk/CRIME.html}{intensitymapping.physics.ox.ac.uk/CRIME.html}}, a software which simulates Stokes Q emission maps for a given frequency. We generate this for our frequency range, and choose the polarisation leakage level of the Stokes Q signal to be 0.5\%. For more details on \texttt{CRIME}, see \citep{Alonso_2014}.

\subsection{\hi cosmological signal}

To generate the \hi cosmological signal, we use the \textsc{MultiDark-Planck} (MDPL2) \citep{Klypin:2014kpa} dark matter N-body simulation, a cosmological simulation which follows 3840$^3$ dark matter particles evolving in a box with side lengths of 1 Gpc/$h$. The cosmology assumed in the simulation is consistent with \textsc{Planck15}, with $\{\Omega_\text{M},\Omega_\text{b}, \Omega_\Lambda, \sigma_8, n_\text{s}, h\} = \{0.307, 0.048, 0.693, 0.823, 0.96, 0.678$\} \citep{Ade:2015xua}. This simulation has been made into the \textsc{MultiDark-Galaxies} \citep{Knebe:2017eei} galaxy catalogue, and the \textsc{MultiDark-SAGE} catalogue was also produced by applying the semi-analytical model \textsc{SAGE} \citep{Croton:2016etl}. These are publicly available in the Skies \& Universe\footnote{\href{http://www.skiesanduniverses.org/page/page-3/page-22/}{skiesanduniverses.org}} web page. We choose to work with the \textsc{MultiDark-SAGE} catalogue.

This simulation is available in the form of redshift snapshots, where each snapshot shows the state of the cosmological density field and galaxies to a different redshift evolution. We choose to use the $z = 0.39$ snapshot, and assume a redshift range of $0.2 < z < 0.58$ and an effective redshift of $z_{\rm eff} = 0.39$. This is not entirely realistic, as it does not include the evolution of the density field and galaxies within this redshift range, but it is sufficient for our purposes. We grid our snapshot of galaxies into voxels using Nearest Grid Point (NGP) assignment.

As first described in \citep{Cunnington:mult}, the method we use for generating \hi intensity mapping simulations is as follows: Each galaxy in the \textsc{MultiDark-SAGE} catalogue has an associated cold gas mass, which we use to compute a \hi mass. For each voxel, we take the \hi mass of every galaxy falling into it, and bin them together. We then convert this binned \hi mass into a \hi brightness temperature $T_{\rm HI}(\boldsymbol{x})$ for that voxel. Doing this for each voxel yields the desired \hi intensity map for our simulation.

One of the limitations of this method is that the \textsc{MultiDark-SAGE} catalogue does not include halos with mass lower than $\lesssim 10^{10}\,h^{-1}$M$_\odot$. In reality, these halos would be present and contain \hi, contributing to the total \hi brightness of each voxel. To account for these missing low mass halos, we rescale the mean \hi temperature of our simulation to a more realistic value. We calculate this realistic value based on the \hi abundance measurement made by the GBT-WiggleZ cross-correlation analysis, $\Omega_{\rm HI} b_{\rm HI} r = [4.3 \pm 1.1] \times 10^{-4}$ \citep{Masui:2012zc}. We assume a cross-correlation coefficient of $r = 1$, and take the \hi bias to be $b_{\rm HI}(z_{\rm eff}) = 1.105$, based on the \hi bias fit from \cite{Villaescusa_Navarro_2018}.

For each redshift slide, this \hi IM simulation can be then transformed into an overtemperature field by subtracting the mean temperature. This overtemperature field traces the underlying dark matter overdensity field, $\delta_{\rm M}(z)$: 
\begin{equation}
    \delta T_{\rm HI}(z) = T_{\rm HI}(z) - \langle T_{\rm HI} \rangle = \langle T_{\rm HI} \rangle b_{\rm HI}(z) \delta_{\rm M}(z)\, .
\end{equation}
\subsection{Noise}\label{noise}

We wish to simulate what the instrumental noise for a MeerKLASS-like survey would look like in our case. The noise is uncorrelated in frequency, and it is Gaussian distributed with a standard deviation of \citep{Alonso_2014}:
\begin{equation}
    \sigma(\nu) = T_{\rm sys}(\nu) \left( \delta_{\nu} t_{\rm tot} \frac{\Omega_{\rm p}}{\Omega_{\rm a} N_{\rm dish}} \right)^{-1/2} \, ,
\end{equation}
where $\delta_{\nu}$ is the frequency resolution of the data (1 MHz in our case), $t_{\rm tot}$ is the total observing time (which we assume to be 1000 hrs), $N_{\rm dish}$ is the number of dishes (64, consistent with MeerKAT), $\Omega_{\rm p} = 1.13 \theta_{\rm FWHM}^2$ is the pixel solid angle, and $\Omega_{\rm a}$ is the survey solid angle:
\begin{equation}
    \Omega_{\rm a} = 4\pi\frac{A_{\rm sky}}{41253} \, ,
\end{equation}
where $A_{\rm sky}$ is the observed sky area (2927 deg$^2$). The frequency dependent system temperature $T_{\rm sys}(\nu)$ is made up of the receiver noise temperature ($T_{\rm rec}$), and the sky temperature ($T_{\rm sky}(\nu)$) \citep{santos2015cosmology}:
\begin{equation}
    T_{\rm sys}(\nu) = T_{\rm sky}(\nu) + T_{\rm rec} \, ,
\end{equation}
where $T_{\rm rec}$ = 25 K and $T_{\rm sky}(\nu)$ is given by:
\begin{equation}
    T_{\rm sky}(\nu) = 1.1 \times 60 \left( \frac{300}{\nu\text{[MHz]}} \right)^{2.55} \, .
\end{equation}

\bsp	
\label{lastpage}
\end{document}